# Pre-Trained Large Language Model Based Remaining Useful Life Transfer Prediction of Bearing

Laifa Tao[a, b, c, d], Zhengduo Zhao [b, c, d], Xuesong Wang [e], Bin Li[b, c, d], Wenchao Zhan[b, c, d], Xuanyuan Su[b, c, d], Shangyu Li[b, c, d], Qixuan Huang[b, c, d], Haifei Liu[b, c, d], Chen Lu[a, b, c, d]; *Zhixuan Lian[a]

[a]*Hangzhou International Innovation Institute, Beihang University, China*
[b]*Institute of Reliability Engineering, Beihang University, China*
[c]*National Key Laboratory of Science and Technology on Reliability and Environmental Engineering, China*
[d]*School of Reliability and Systems Engineering, Beihang University, China*
[e]*China Electronics Information Industry Group Co., Ltd. Sixth Research Institute, Beijing, China*

*Corresponding Author: Zhixuan Lian*

E-mail: taolaifa@buaa.edu.cn; zhengduozhao@buaa.edu.cn; 99402006@sina.com; libin1106@buaa.edu.cn; 20376049@buaa.edu.cn; suxuanyuan@buaa.edu.cn; lishangyu@buaa.edu.cn; qxhuang@buaa.edu.cn; phoebeliu@buaa.edu.cn; luchen@buaa.edu.cn; * lianzhixuan@buaa.edu.cn;

**Abstract:** Accurately predicting the remaining useful life (RUL) of rotating machinery, such as bearings, is crucial for equipment reliability and minimizing unexpected failures in industrial systems. Despite recent advancements, data-driven deep learning methods face challenges in practical industrial settings due to inconsistent data distributions between training and testing phases, and limited generalization capabilities for long-term RUL predictions. To address these issues, we propose LM4RUL, a framework for RUL prediction based on pre-trained Large language Model (LLM). LM4RUL leverages the generalization and reasoning capabilities of LLM to transfer predictive knowledge from pre-training, effectively overcoming data inconsistencies and enhancing prediction accuracy. This represents a meaningful advancement in the artificial intelligence field, being among the first efforts to successfully apply LLM to RUL prediction tasks without the need for additional manual instruction, thereby extending the boundaries of AI applications beyond natural language processing and into complex industrial scenarios. The framework includes the local scale perception representation component, which captures fine-grained bearing degradation trends by tokenizing vibration data, and hybrid embedding learning, which selectively freezes and fine-tunes parameters to model complex nonlinear degradation. A two-stage fine-tuning approach further integrates pre-trained knowledge, enhancing adaptability without complex architectures. Experiments on authoritative datasets demonstrate that LM4RUL significantly improves long-term RUL prediction and adaptability, achieving high precision and robustness. This work bridges a critical gap in RUL prediction with advanced AI techniques and highlights its practical impact on equipment reliability and operational safety, showcasing its value for predictive maintenance in industrial AI applications.

## 1. Introduction

Bearings are critical in industrial systems and prone to failures under high-speed and heavy-load conditions, leading to transmission system paralysis, safety accidents, and economic losses (Cui et al., 2024; Lu et al., 2024; Xiao et al., 2025). Prognostics and Health Management (PHM) technology monitors operational parameters of bearings through sensors, uses intelligent algorithms to identify their status, and implements condition-based maintenance. It is widely applied in various industries, reducing maintenance costs and improving equipment reliability and safety (Sun & Wang, 2024; Zhu et al., 2023a). Among these technologies, RUL prediction is the core of the PHM system (Magadán et al., 2024). Accurate RUL prediction assesses the degradation state of bearings, prevents catastrophic accidents, and detects changes in bearing conditions earlier through long-term extrapolation, thus being of great importance. However, long-term RUL prediction is highly challenging.

Researchers have developed various RUL prediction algorithms and models, mainly classified into model-based methods, data-driven methods, and hybrid methods. Model-based methods use historical data to develop mathematical or physical models to describe equipment degradation patterns accurately. For example, Lei et al. proposed a two-stage mechanical RUL prediction method (Lei, Li, Gontarz, et al., 2016). In the first stage, they constructed a health indicator integrating multiple features; in the second, they combined maximum likelihood estimation and particle filtering algorithms to estimate RUL. Hu et al. modeled the relationship between temperature parameters of wind turbine bearings and RUL based on the Wiener process (Hu et al., 2018), considering uncertainties from wind speed and direction. Combining the failure principle of temperature parameters exceeding the warning threshold, they established an RUL prediction model for wind turbine bearings using the inverse Gaussian distribution. Theoretically, accurate physical or mathematical models for RUL prediction are ideal; however, these model-based methods are difficult to develop accurately due to the complexity of degradation mechanisms. Additionally, literature on model-based methods has focused on constructing health indicators (HI) based on specific empirical knowledge for decades. These limitations restrict the practical application of model-driven methods due to the need for in-depth analysis of failure mechanisms and extensive prior knowledge.

Hybrid methods combine data-driven and model-based methods, enhancing data-driven models' performance using physical knowledge, thus reducing dependence on extensive expert knowledge. For example, Deng et al. proposed a hybrid transfer learning architecture based on calibration (Deng et al., 2023), considering crack and spalling behavior in the bearing degradation process. They designed a physical information Bayesian network, using calibration information from the physical model as an enhanced input space for the data-driven model. However, this hybrid method is still knowledge-intensive and requires a deep understanding of degradation mechanisms. Additionally, due to the high complexity of simulations in physical models (Jiang et al., 2025), the computational overhead is large (Ferreira & Gonçalves, 2022). Therefore, whether using physical theory-based or hybrid methods, dealing with complex industrial systems requires extensive expertise and computational resources, with high costs limiting widespread adoption and implementation.

Unlike the first two methods, data-driven methods rely on monitoring data containing degradation information from equipment or systems rather than expert knowledge (Jiang et al., 2025). These methods use artificial intelligence algorithms to automatically learn and extract patterns, making them suitable for analyzing large amounts of data collected by the Industrial Internet of Things (IIoT) (H. Zhao et al., 2021). For example, Li et al. explored time-frequency domain features in bearing RUL prediction, proposing a method for multi-scale feature extraction using convolutional neural networks and validating it on a publicly available rolling bearing dataset (Li et al., 2019). Zhu et al. used wavelet transform to extract time-frequency domain features and constructed a multi-scale convolutional neural network model to estimate RUL (Zhu et al., 2019), emphasizing that extracted features should preserve both global and local information. To address prediction challenges from the high nonlinearity and complexity of mechanical systems, Cao et al. proposed a deep learning framework combining residual attention mechanism and spatiotemporal convolutional networks (Cao et al., 2021), inputting the marginal spectrum of bearing vibration signals to achieve RUL prediction. Wang et al. proposed a deep learning RUL prediction model called deep separable convolutional network (Wang et al., 2019), considering correlations between different sensor data, which can automatically learn high-level representations directly from vibration signals to achieve end-to-end RUL prediction. Ma et al. conducted architectural fusion based on convolutional neural networks and long short-term memory networks (M. Ma & Mao, 2021), deeply mining degradation information through stacked encoding-prediction architecture. With the advantages of minimal manual labor, little prior knowledge, and ease of implementation, data-driven RUL prediction methods have seen significant development. However, data-driven bearing RUL prediction methods are still constrained by three key problems.

**The gap between ideal data conditions and actual industrial scenarios becomes more pronounced over the long term RUL.** Specifically, the ideal assumption that "training data and test data are independently and identically distributed" is difficult to meet in actual industrial scenarios, mainly due to two factors: First, during the degradation process, the state of bearings changes continuously and is affected by external interference. The distribution of degradation information drifts under time-varying characteristics (Liu & Li, 2024; Ochella et al., 2022), Even bearings of the same model under the same operating conditions have different degradation feature distributions, with more significant differences under different operating conditions(Kumar et al., 2024). Second, it is almost impossible for the training data to cover all operating conditions, as the cost of collecting complete operating condition and full lifecycle data in an experimental environment is nearly unacceptable (Shi et al., 2024). More realistic is having run-to-failure data from similar target domains under different operating conditions and a small amount of degradation data under target conditions (Guo et al., 2024). Using the XJTU-SY dataset as an example, we extracted the Root Mean Square (RMS) features of bearings with the same failure behavior under three operating conditions and presented them as probability density distributions. As shown in Figure 1, the horizontal axis represents RMS, and the vertical axis represents the probability distribution. Even under the same operating conditions and failure behavior, degradation characteristic

distributions differ, and this difference is greater under different conditions. Current mainstream research rarely considers the impact of these distribution changes.

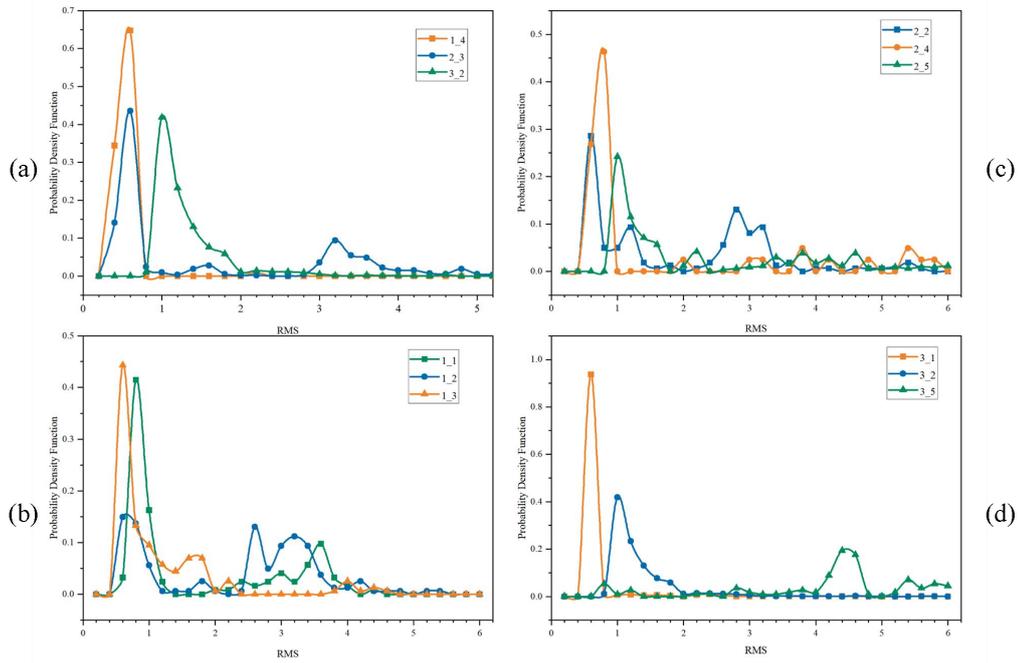

Figure 1 Probability density distribution of RMS characteristics of bearings under different operating conditions

**Existing methods have limitations in generalization for long-term RUL prediction.** Many studies confirm that transfer learning technology can overcome data distribution differences caused by mechanical damage, operational condition changes, and external interference (Qin et al., 2024). Typical methods include discrepancy metrics (Mao et al., 2020) and domain adversarial training (Ganin et al., 2016). These transfer strategies reduce differences between the source and target domains in the feature space, assuming the source domain data provides sufficient generalization information and the model can fully learn key degradation information(Xiang et al., 2020). However, in practice, these transfer strategies often fail to learn complete domain common information from training data, resulting in learned domain common features being a limited subset. These transfer strategies allow the model to extract limited generalized features, making it difficult to learn helpful contextual information for long-term RUL prediction. This explains why most studies are limited to single-step RUL prediction rather than long-term prediction. Fortunately, the pre-training-fine-tuning transfer learning strategy offers another perspective (Ji et al., 2023; Li et al., 2024). This method aims to utilize all knowledge accumulated during the pre-training phase, rather than just learning some generalization features, and then transfer it to the target task, performing tasks under various distributed data. In natural language processing (NLP) research, pre-trained large models can predict tokens over a long window (W. X. Zhao et al., 2023). However, in RUL prediction, long-term RUL prediction methods based on the pre-training-fine-tuning paradigm have not been fully studied.

**The limitations of "small model" architectures in embedding learning need emphasis.** "Small models" typically refer to models with fewer parameters and relatively simple feature extraction architectures, such as convolutional architectures in (Long et al., 2015) and recurrent neural architectures in (Zaremba et al., 2015). "Small models" have limitations in data-driven RUL prediction (Zhang et al., 2023), mainly in limited representation ability and high design complexity in transfer scenarios. Although CNNs can directly extract sensitive change features, their local receptive fields cannot capture global features effectively, and they are insufficient in capturing long-term dependencies, requiring recurrent units to process vibration data. While RNNs are suitable for modeling temporal dependencies, their instability in gradient propagation during optimization makes learning long-term degradation features challenging. Additionally, network architectures in existing studies, combined with aforementioned transfer strategies, require fine-tuning the loss function design or complex gradient propagation of multiple adversarial network modules. They struggle to extract intrinsic properties of degradation features from multiple scales, such as global and local, and relate them to macro degradation trends.

To address challenges in data-driven RUL prediction, such as data distribution differences in transfer scenarios, insufficient long-term prediction performance, and feature learning limitations of smaller models, this paper introduces LM4RUL, a novel framework for bearing RUL prediction based on pre-trained large language models (LLMs). Leveraging the predictive power and deep feature learning of LLMs, LM4RUL captures long-term contextual information and addresses distributional shifts between training and testing data. The framework comprises two key components: the Local Scale Perception Representation (LSPR) component, which extracts local degradation trends, and the Hybrid Embedding Learning component based on a Frozen Pre-trained Large Model (FPLM-HEL), enhancing long-term degradation feature learning in transfer scenarios. First, vibration signals are truncated through FPT detection and processed using the LSPR component, which applies Short-Time Fourier Transform (STFT) to extract time-frequency features, normalizes these features by channel, and tokenizes them through patching to encapsulate local degradation information. This allows the large model to learn degradation trends over extended historical windows at fine granularity. The FPLM-HEL component selectively fine-tunes a small portion of the pre-trained LLM while freezing most parameters, retaining general predictive knowledge to capture the intrinsic rules of bearing degradation. Ultimately, the model outputs RUL predictions based on this layered representation of degradation behavior. In contrast to traditional RUL approaches, which rely heavily on handcrafted feature engineering and require substantial customization to adapt to differing environments, LM4RUL provides a more adaptive, automated, and transferable solution for accurate long-term RUL estimation. The contributions of this paper can be summarized as follows:

**1. LM4RUL as a Novel Bearing RUL Prediction Framework:** We propose LM4RUL, a novel transfer learning framework that leverages pre-trained large language models (LLMs) for bearing RUL prediction. Unlike traditional methods, LM4RUL avoids the need for extensive data collection from scratch and utilizes pre-training strategies to achieve long-term RUL extrapolation with superior generalization capabilities. To our knowledge, this is the first application of LLMs in bearing RUL prediction, marking a significant advancement in predictive maintenance.

**2. LSPR Component as a Translator for Bearing Degradation to LLMs:** The LSPR component enhances adaptability to bearing degradation sequences by tokenizing raw vibration data and capturing fine-grained degradation trends. This enables LLMs to effectively process and represent local degradation features, aligning well with the Transformer-based architecture used in LM4RUL. LSPR bridges the gap between LLM language understanding and bearing degradation data, acting as a compiler that translates degradation information into a comprehensible form for LLMs.

**3. FPLM-HEL Component for Enhanced Feature Extraction and Learning:** FPLM-HEL component selectively freezes and fine-tunes pre-trained model layers, retaining predictive knowledge while effectively modeling the complex nonlinear relationships inherent in degradation processes. FPLM-HEL employs a triple-embedding approach that enhances the model's ability to extract and learn abstract degradation features, contributing to precise long-term RUL prediction. By leveraging selective fine-tuning, FPLM-HEL inherits the strengths of Transformer architectures, effectively capturing intricate degradation patterns.

**4. Two-Stage Fine-Tuning for Efficient Generalization:** We propose a two-stage fine-tuning strategy that activates the generalization capabilities of LM4RUL without the need for complex transfer learning architectures. This method efficiently adapts the pre-trained model to diverse industrial scenarios, reducing the data requirements typically needed to train RUL prediction models from scratch.

**5. Experimental Validation Demonstrating Superiority:** Comprehensive experiments on authoritative datasets (e.g., XJTU-SY and FEMTO) demonstrate that LM4RUL outperforms conventional models in both prediction accuracy and adaptability. The analysis reveals an intrinsic connection between the feature extraction behavior of LM4RUL and Principal Component Analysis (PCA), highlighting LM4RUL's ability to identify key degradation features that enhance generalization in RUL prediction tasks. These results validate the feasibility of using LLMs for effective RUL prediction in practical industrial applications, underscoring both the accuracy and efficiency of the proposed framework.

## 2. Related Work

### 2.1 Long-Term RUL Prediction with Cross-Condition Generalization

To achieve accurate long-term RUL prediction for bearings, overcoming the challenge of across-condition generalization is key (Zhu et al., 2023b). Existing models, such as Yan et al.'s On-LSTM, combine frequency domain features with an LSTM-tree structure to improve predictive accuracy (Yan et al., 2020). However, these models struggle with long-term across-condition generalization, especially under varying operating conditions. Similarly, Chen et al.'s QFMDCAET framework adapts health indicators to different operating conditions using multi-scale deep learning methods, but its reliance on complex feature extraction and manual design hinders scalability in long-term prediction scenarios (D. Chen et al., 2023a). Qin et al.'s MLMA-Net enhances memory capacity using attention layers, but is still susceptible to instability when faced with diverse temporal data distributions (D. Chen et al., 2023b). These models focus on capturing temporal dependencies and historical degradation features but fall short in handling long-term temporal generalization effectively.

## 2.2 Transformer-based Large Language Models

The successful application of Transformer-based LLMs across fields demonstrates their powerful capabilities (Zhao et al., 2023), owing primarily to two advantages: 1) LLMs can predict long-window token sequences, producing optimal outputs based on historical input distributions; and 2) they exhibit strong domain generalization, achieving effective transfer prediction in few-shot and zero-shot scenarios. Through pre-training, LLMs acquire general predictive knowledge, enhancing generalization across diverse word vector distributions. Leveraging these strengths, recent studies have applied LLMs for time series prediction in finance (Zhang et al., 2023), healthcare (Li et al., 2024), and climate (Nguyen et al., 2023). For instance, Sun et al. proposed the Text Prototype Aligned Embeddings (TEST), which aligns time series embeddings with the LLM input space using soft prompting to enhance learning, achieving results on par with or superior to current SOTA models (Sun et al., 2023). Chang et al. introduced LLM4TS, aligning LLM features with time series data through calibration and fine-tuning with LoRA, outperforming SOTA methods (Chang et al., 2024). Similarly, Zhou et al. developed OFA, a time series framework based on GPT-2 that achieves or exceeds SOTA performance through fine-tuning limited parameters (Zhou et al., 2023). N-gram theory suggests that LLMs possess inherent predictive capabilities (Brown et al., 1992), highlighting their promise for time series transfer prediction, a critical insight for bearing RUL prediction involving multi-sensor time series regression. Compared to smaller models, LLMs utilize multi-layered structures with extensive parameters, enabling few-shot and zero-shot adaptability, demonstrating stronger knowledge retention and generalization reasoning capabilities(Pang et al., 2024). Studies like TIMELLM (Jin et al., 2024), OFA (Zhou et al., 2023), and LLM4TS (Chang et al., 2024) confirm these models outperform SOTA in cross-modal time series tasks with only task-specific fine-tuning. In RUL prediction, which entails multivariate time series regression to model nonlinear relationships between degradation features and RUL (Lei et al., 2018), LLMs offer a compelling solution to key challenges. First, effective methods are needed to process high-frequency, noisy bearing signals—requiring specific tokenization strategies compatible with Transformer architectures. Second, current model architectures must be adapted to handle multi-channel degradation data, moving beyond the text-based outputs typical of LLMs to map degradation patterns to RUL values. Finally, fine-tuning strategies are necessary to utilize LLMs' vast parameter scales without full-scale retraining, which is impractical for RUL prediction.

To address the challenge in RUL prediction, we propose targeted solutions by integrating LLMs: 1) Data processing is improved by our LSPR component, which directly utilizes time-frequency domain features from vibration signals, enabling the model to learn both macro degradation trends and fine-grained key degradation patterns without complex manual HI construction; 2) In the model architecture, the FPLM-HEL component selectively freezes and fine-tunes specific layers, adding neural mapping layers to retain pre-trained general predictive knowledge, and effectively modeling the complex nonlinear relationship between high-dimensional degradation and RUL sequences; 3) For training strategy, we introduce a two-stage fine-tuning method that activates the domain adaptation capabilities of FPLM-HEL while avoiding the computational cost of pre-training a large model from scratch, enabling efficient transfer of knowledge for bearing RUL prediction.

## 3. Methodology

The proposed life prediction framework mainly includes four steps, LSPR covers the main operations of the first and second steps, FPLM-HEL covers all operations of the third step, as detailed in Figure 2: 1) First is the acquisition of raw vibration samples. After the bearing vibration signals are collected by sensors, FPT detection is performed to capture vibration data with significant degradation information, which are then segmented into multiple vibration samples through fixed windows. 2) Data preprocessing and feature extraction. The obtained vibration signal samples are divided into two-stage fine-tuning sample sets, the supervised fine-tuning set and the prompt fine-tuning set. For the vibration data in each fine-tuning sample set, the STFT method is used to extract time-frequency domain features for subsequent model input. 3) Degradation sequence embedding learning. The LM4RUL model is constructed to model the complex nonlinear relationship between time-frequency features and RUL values based on the two-stage fine-tuning method. 4) The data obtained from online monitoring are input into our framework to achieve RUL prediction. The specific details are as follows.

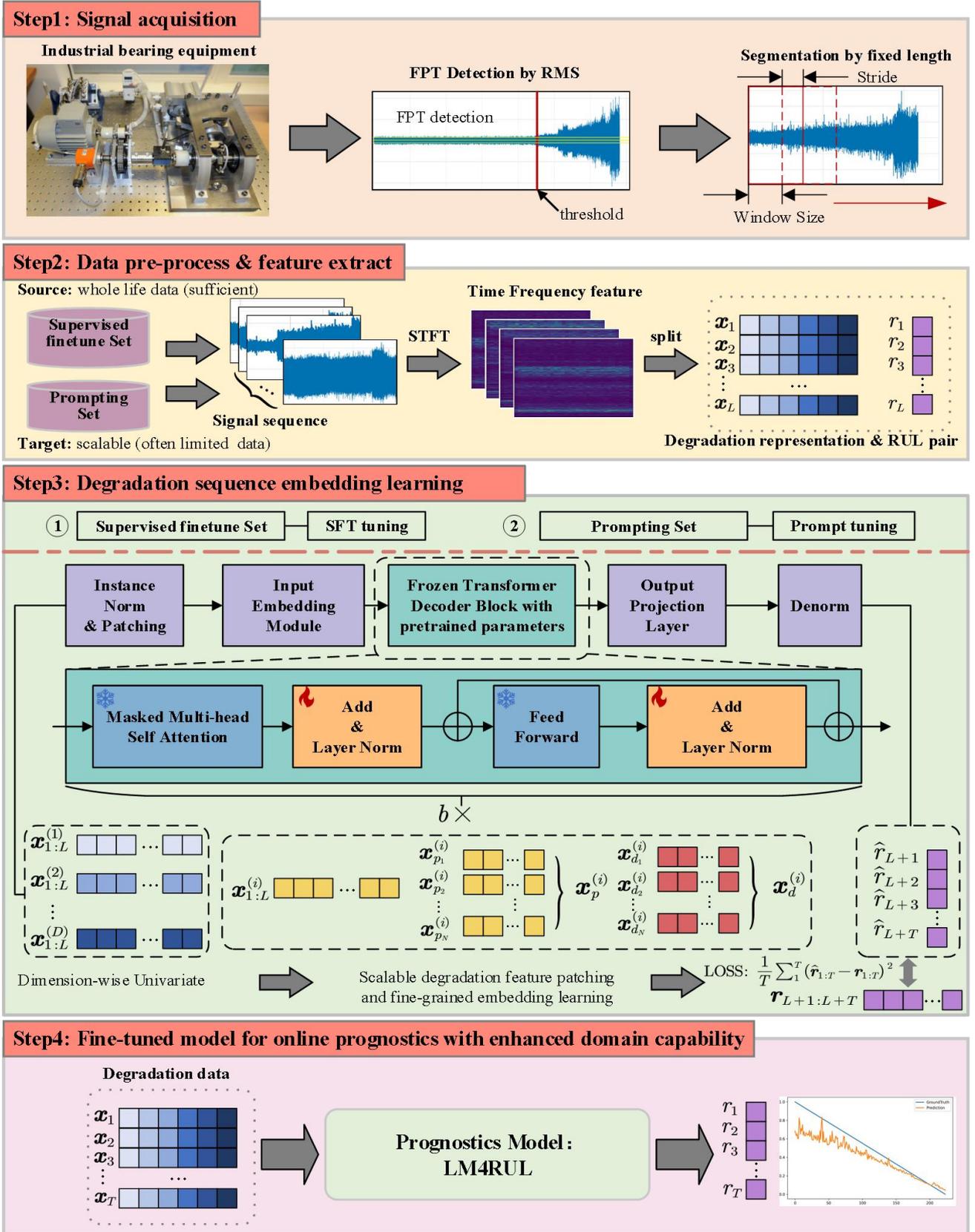

Figure 2 The overall process of the proposed framework

## 3.1 Problem Formulation

We start with formally introducing the problem of bearing long-term RUL prediction. RUL prediction is essentially a regression task of predicting a univariate time series from a multivariate time series. Given a multivariate time series $x_{1:L} = [x_1, x_2, ..., x_L]$ with a lookback window length of $L$, the goal is to predict the future univariate time series $r_{L+1:L+T} = [r_{L+1}, ..., r_{L+T}]$ for $T$ steps ahead. Each vector $x_t \in \mathbb{R}^D$ at time step $t$ contains the degradation information of the bearing corresponding to this timestamp, while the value $r_t \in \mathbb{R}^1$ is the RUL value. Ideally, the most desirable RUL prediction is extrapolative, meaning that the value of $T$ should cover multiple time steps, i.e., $T \geq 2$. When the prediction error is within a reasonable acceptance range, the larger the value of $T$, the more future trends in remaining useful life can be estimated from early degradation monitoring signals, thus making predictive maintenance truly valuable for practical engineering use.

### 3.2 LSPR Component

The LSPR component consists of adaptive FPT detection to slice out data containing the most degradation information, extraction of time-frequency domain features using STFT, and a patching operation to obtain subsequent token sequences by aggregating key degradation information at the local scale for input to the FPLM. LSPR matches the raw vibration signal data of the bearing to the input pattern of the LLM. Details about LSPR component is shown as Figure 3.

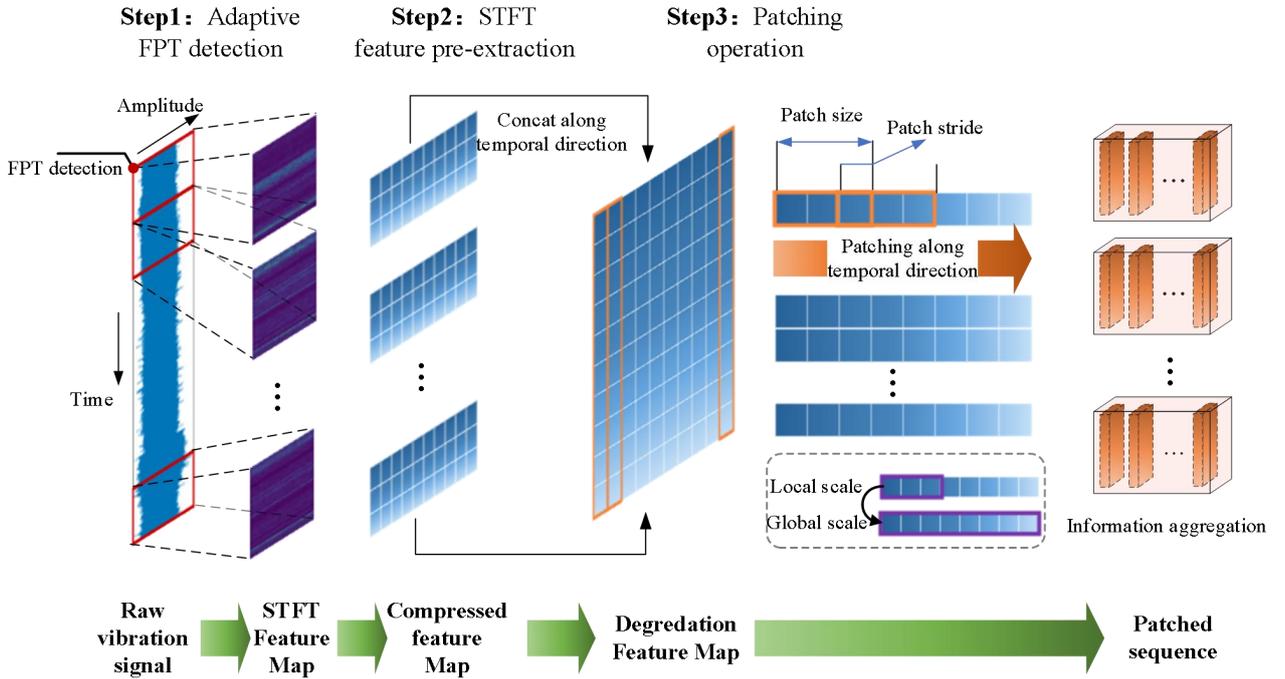

Figure 3 The architecture of the proposed LSPR components

#### 3.2.1 FPT Determination

FPT is the first step in our RUL prediction framework, which determines the time point when the bearing starts to show significant degradation to delimit the data range for training the model. Bearing degradation generally has two stages: the normal operation stage and the failure stage. Since early bearing failures mostly occur randomly, the failure characteristics are weak and difficult to detect. The signal-to-noise ratio of the vibration signals in this stage is relatively low, containing less degradation information and more interfering noise. If the initial data is input into the model for training, it often makes it difficult for the model to effectively utilize the data, resulting in incorrect RUL estimation. On the other hand, if the FPT selection is unreasonably delayed, key information is excluded from the prediction process, which also reduces the accuracy of RUL prediction. Therefore, choosing an appropriate FPT can maximize the exclusion of interfering noise and retain degradation information, which is important for subsequent RUL prediction. The Root Mean Square (RMS) indicator is widely used in RUL prediction because RMS empirically increases with the severity of component degradation. Therefore, this study uses this recognized indicator for FPT detection. That is, when the RMS reaches a predetermined threshold, the FPT point is detected at this time.

FPT is detected by calculating the $3\sigma$ interval threshold of RMS, formalized as:

$$cv_t = \mu_{t-1} \pm 3\sigma_{t-1}$$
$$= \frac{1}{t-1}\sum_{i<t} RMS_i \pm 3\sqrt{\frac{1}{t-1}\sum_{i<t}(RMS_i - \mu_{t-1})^2} \quad (1)$$

where $\mu_{t-1} = \frac{1}{t-1}\sum_{i<t} RMS_i$ represents the mean value of RMS, $\sigma_{t-1} = \sqrt{\frac{1}{t-1}\sum_{i<t}(RMS_i - \mu_{t-1})^2}$ represents the variance.

If the RMS value exceeds the threshold, it indicates that the prediction process needs to be triggered. Since during the normal operation stage, the presence of random noise is unlikely to cause multiple consecutive RMS values to exceed the 3-sigma range, Lei et al. used a consecutive trigger mechanism to avoid the impact of random errors (Lei, Li, & Lin, 2016). This study follows this setup, with the consecutive trigger mechanism process described as follows: let three consecutive RMS values form a group, If the condition $\{RMS_{t+1}, RMS_{t+2} > cv_t \mid RMS_t > cv_t\}$ is met, that is, all three consecutive RMS values exceed $cv_t$, the time $t$ is considered to correspond to the initial FPT value; otherwise, continue traversing $t$ until the condition is met.

By reasonably selecting FPT in the above manner, the maximum amount of useful degradation information can be retained for subsequent RUL prediction.

### 3.2.2 Feature pre-extraction based on Short Time Fourier Transform

Since firstly introduced, STFT is applied widely in various tasks to process nonstationary signals with explicit physical meaning. It can be denoted as：

$$X(t,f) = \int_{-\infty}^{+\infty} x(\tau)w(\tau-t)e^{-i2\pi f\tau}d\tau \quad (2)$$

where $x(t)$ is signal, $w(\tau-t)$ is the window function, $f$ represents frequency.

$X(t,f)$ is considered to be the time-frequency domain features of the input signal. In this paper, we use the energy distribution $E(t,f)$ of the raw signal (i.e., the spectrogram) as subsequent input, which can be formulated as:

$$E(t,f) = |X(t,f)|^2 \tag{3}$$

where $E(t,f)$ is computed by the results of STFT operation $X(t,f)$.

Let $x_{FPT}(t)$ be the raw signal after FPT determination, then applying a fixed length window on it to obtain multiple sub-sequences. Each subsequence will be transformed into time-frequency feature map $f_{m \times n}$ of size $m \times n$, where $m$ denotes the dimension of frequency axis and $n$ denotes the dimension of time axis. Then concatenate all feature map along the time direction to obtain degradation feature map $F_{L \times D} = [f^0, f^1, ..., f^N]^T$, where $N$ is the number of all sub-feature maps, $L = N \times n$ denotes the temporal dimension of degradation feature map, and $D$ is the dimension of feature.

### 3.2.3 Degradation Sequence Representation Based on Channel-Independence Patching

The goal of RUL prediction models is to effectively learn the correlations between data at different time steps within a sequence. However, a single time point in historical data does not carry in-context information like a word in a sentence, and learning locally aggregated information is more helpful in understanding the long-term trend of bearing degradation. Therefore, in RUL sequence prediction, it is necessary to explore the local degradation scale of input data in a more effective manner. Specifically, based on Patching operations (Nie et al., 2023), we enhance locality by aggregating time steps into sub-series level patches to capture higher-level degradation information that is unattainable at single time point levels. The specific details are as follows:

For the ease of illustration, given a degradation feature map denote as $F: [x_{1:L}^{(1)}, x_{1:L}^{(2)}, ..., x_{1:L}^{(i)}, ..., x_{1:L}^{(D)}] \in \mathbb{R}^{L \times D}$. The $i$-th univariate feature sequence from time index 1 to $L$ is represented as $x_{1:L}^{(i)} \in \mathbb{R}^L$. Each univariate feature sequence is then segmented into overlapping or non-overlapping patches. Let the patch size be $P$, and the stride between patches be $S$, Each $x_{1:L}^{(i)}$ is segmented into $N$ patches of length $P$, such that $N = \frac{L-P}{S} + 2$. Thus, each $x_{1:L}^{(i)}$, after the patching operation, is transformed into a series of patches that perceive local degradation trends. The patch matrix $x_p^{(i)} \in \mathbb{R}^{N \times P}$ corresponding to each univariate feature sequence is represented as: $x_p^{(i)} = (x_{p1}^{(i)}, x_{p2}^{(i)}, ..., x_p^{(i)})$.

It is worth noting that for sequences whose lengths cannot be evenly divided into patches of equal length, we use a padding operation at the end of each original sequence $x_{1:L}^{(i)}$. The last value of $x_{1:L}^{(i)}$ is repeated $S$ times and added to the end of the original sequence to ensure that each patch maintains a length of $P$. This prevents the last patch from being discarded due to insufficient length, ensuring that no critical degradation information is lost.

By using the patching operation, the degradation feature map $F$ is reorganized from size $L \times D$ to size $D \times N \times P$, allowing the model to effectively focus on local-scale degradation features, associate them with the global degradation trend, and accommodate longer historical sequences. This enables the model to learn the degradation features at a granularity that includes richer trend information.

## 3.3 RUL Modeling Based on FPLM-HEL Components

The detailed model architecture of the FPLM-HEL component is illustrated in Figure 4, where the upper part shows the network structure and the lower part shows the freezing and fine-tuning of the parameters of each layer during the fine-tuning stage in different colors with respect to the relative position of each layer. The FPLM-HEL component is mainly composed of input embedding module, FPLM module, RIN layer and output projection layer. Among them, the input embedding module consists of three parallel encoders executing different functions, and the degradation representation sequence after patching is fed into the input embedding module to encode the feature information, absolute position information, and relative temporal information to obtain the embedding and align with the feature space of the LLM. Then the FPLM module freezes and fine-tunes different sub-layers in a two-stage fine-tuning, passes on the general predictive knowledge from the pre-training, and adapts stage by stage to the distributional properties of the target domain data, characterizing the input embedding in depth after passing through stacked multiple FPLM blocks. The hidden layer state passes through the projection layer and finally outputs the estimate RUL for a future period.

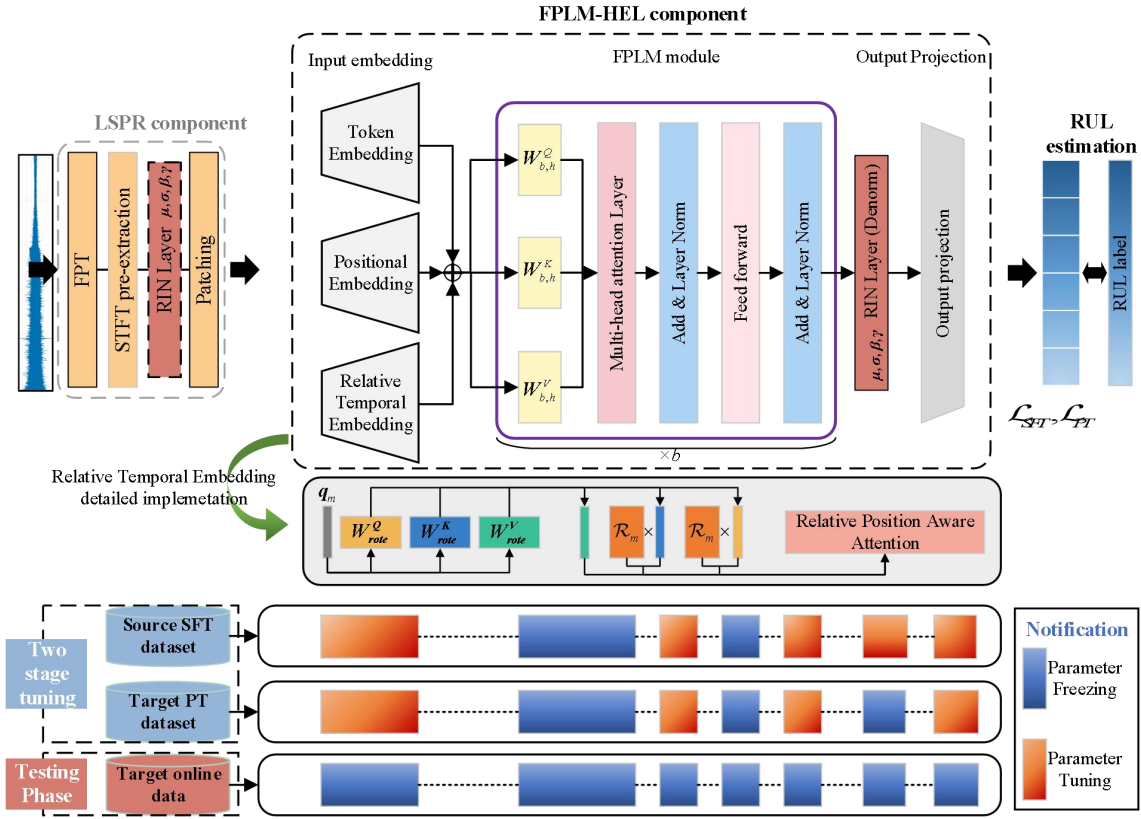

Figure 4 The architecture of the proposed FPLM-HEL components

### 3.3.1 Reversable Instance Normalization Layer based on Learnable Affine transformation

As discussed in introduction, the bearing degradation characteristics vary greatly under different operating conditions and do not satisfy the ideal condition of training and testing independently and identically distributed. In order to remove the non-stationary discrepancy information at the data input stage, we use a set of learnable affine transform parameters in the progress of the normalization of the features extracted from the STFT (X. Ma et al., 2024), These affine transformation parameters form a reversible instance normalization layer with a symmetric structure that removes non-stationary information from the degradation characteristics of bearings under different operating conditions. The details of the RIN are as follows.

RIN consists of normalization and denormalization layers with a symmetric structure. First, standardization is performed by calculating the mean and variance of each instance, we illustrate an example of a univariate case. Let the univariate sequence of a certain channel in the multivariate time series be represented as $x_{1:L}^{(i)}$, we calculate the mean and variance of $x_{1:L}^{(i)}$ from each instance $x_j^{(i)} \in \mathbb{R}^1, j \in [1, L]$, and the calculation formulas are as follows:

$$\mathbb{E}\left[x_{1:L}^{(i)}\right] = \frac{1}{L}\sum_{j=1}^{L} x_j^{(i)}$$
$$\mathrm{Var}\left[x_{1:L}^{(i)}\right] = \frac{1}{L}\sum_{j=1}^{L}\left(x_j^{(i)} - \mathbb{E}\left[x_j^{(i)}\right]\right)^2 \qquad (4)$$

Next, $x_{1:L}^{(i)}$ is normalized using a set of learnable affine transformation parameters, represented as:

$$\hat{x}_j^{(i)} = \gamma_j \times \left(\frac{x_j^{(i)} - \mathbb{E}\left[x_{1:L}^{(i)}\right]}{\sqrt{\mathrm{Var}\left[x_{1:L}^{(i)}\right] + \epsilon}}\right) + \beta_j \qquad (5)$$

where $\gamma_j, \beta_j \in \mathbb{R}^1$ represents the learnable affine transformation parameters, $\epsilon$ denotes very small numbers avoiding a denominator of 0. Assuming that the final output of the model is the sequence $\tilde{y}^{(i)}$, and the inverse normalized output is $\hat{y}^{(i)}$, for each instance denoted as:

$$\hat{y}_j^{(i)} = \sqrt{\mathrm{Var}\left[x_{1:L}^{(i)}\right] + \epsilon} \times \left(\frac{\tilde{y}_j^{(i)} - \beta_j}{\gamma_j}\right) + \mathbb{E}\left[x_{1:L}^{(i)}\right] \qquad (6)$$

When there is a drift in the distribution of the training data from the test data, it can be denoted as $P_{train}(x) \neq P_{test}(x)$, hence $\mathbb{E}[x_{train}] \neq \mathbb{E}[x_{test}]$ as well as $Var[x_{train}] \neq Var[x_{test}]$ holds. From Equation (5), for $\forall i \in [1, D]$, $\mathbb{E}[\hat{x}_{1:L}^{(i)}] = \boldsymbol{\beta}$ and $\mathrm{Var}[\hat{x}_{1:L}^{(1)}] = \boldsymbol{\gamma}^2$ hold, where $\boldsymbol{\beta} = [\beta_j]_{j=1}^L$, $\boldsymbol{\gamma} = [\gamma_j]_{j=1}^L$. Symmetrically, since the affine transform parameter layer is frozen during model testing, i.e., for both $x_{1:L}^{(i)} \sim P_{train}$ and $x_{1:L}^{(i)} \sim P_{test}$, the mean and variance are the same. From Equation (6),

its vector form is:

$$\hat{\boldsymbol{y}}^{(i)} = \frac{\tilde{\boldsymbol{y}}^{(i)} - \boldsymbol{\beta}}{\boldsymbol{\gamma}} \cdot \sqrt{\text{Var}\left[\boldsymbol{x}_{1:L}^{(i)}\right]} + \mathbb{E}\left[\boldsymbol{x}_{1:L}^{(i)}\right] \tag{7}$$

therefore the calculation of mean and variance for it can be expressed as $\mathbb{E}[\hat{\boldsymbol{y}}^{(i)}] = \Delta + \mathbb{E}[\boldsymbol{x}_{1:L}^{(i)}]$ and $\text{Var}[\hat{\boldsymbol{y}}^{(i)}] = \lambda \cdot \text{Var}[\boldsymbol{x}_{1:L}^{(i)}]$, where $\Delta, \lambda$ denotes a statistical difference between training and testing that is not relevant for the input data model only. Therefore, RIN actively removes the non-stationary statistics that characterize the degradation of bearings from different operating conditions. Similar methods such as Layer Norm (Xu et al., 2019) and Batch Norm (Ioffe & Szegedy, 2015) are widely used in Transformer backbone networks, which have been extensively researched and proven to stabilize gradient descent during training and reduce the risk of overfitting.

### 3.3.2 Input Embedding Module

In this subsection, we propose the Input Embedding Module to align degradation patch sequences with the LLM's feature space. In this process, local sensitive features and global degraded features are fully extracted, with absolute position and relative temporal position information encoded and embedded.

1) Token embedding encoder

Token embedding encoder further extracts features in patch sequence that are sensitive to variations on local scale, thus for the patched degradation representation sequence, it is first passed through a convolutional layer to map $\boldsymbol{x}_p^{(i)} \in \mathbb{R}^{N \times P}$ to the latent space with the hidden layer dimension $d$ of the FPLM, resulting in $\boldsymbol{e}_d^{(i)}$, Additionally, an additive learnable positional encoding denote as to monitor the absolute temporal order, which can be represented as:

$$\boldsymbol{e}_{token}^{(i)} = \text{Layer}_{conv}(\boldsymbol{x}_p^{(i)}) \tag{8}$$

2) Positional embedding encoder

Positional embedding aims to encode absolute position information. Since the multi-head attention mechanism in the FPLM module does not have the ability to actively encode the absolute positional relationships of tokens, it needs to be added explicitly by hands. There are multiple encodings that have been developed to track the absolute position of each token, and we use simple and efficient sine-cosine positional encodings instead of a learnable parameterized encoding layer to reduce the burden of unnecessary parameter learning. Position embedding formulas are as follows:

$$\boldsymbol{e}_{pos(n,j)}^{(i)} = \begin{cases} E_{pos}(n, 2j) = \sin(n / 10000^{2j/d}) \\ E_{pos}(n, 2j+1) = \cos(n / 10000^{2j/d}) \end{cases} \tag{9}$$

where $n$ is the absolute position of each token in patch sequence, index $j \in [0, d/2-1]$ distinguishes between odd and even positions, $d$ is the hidden dimension of FPLM blocks.

3) Relative temporal embedding encoder

To enable the FPLM module to perceive relative temporal information in the patch sequence, we perform relative position-aware attention on the tokens. This involves introducing relative temporal information while extracting global correlation features in the attention layer, achieved by using a rotational encoding matrix. Specifically, we propose relative position-aware attention based on standard scaled dot-product attention. The calculation procedure is shown below.

Let the query matrix in relative position-aware attention be $W_{rote}^{Q} \in \mathbb{R}^{N\times d}$, the key matrix be $W_{rote}^{Q} \in \mathbb{R}^{N\times d}$, value matrix be $W_{rote}^{V} \in \mathbb{R}^{N\times d}$, where $N$ is the length of the patch sequence and $d$ is the hidden layer dimension in FPLM. For ease of illustration, for the $i_{th}$ patch sequence $x_{p}^{(i)} \in \mathbb{R}^{N\times P}$, where the token at any two positions can be represented as patch $x_{pm}^{(i)}$ at position $m$ and patch $x_{pn}^{(i)}$ at position $n$, can be mapped by the query matrix and the key matrix as a query vector $q_m, q_n \in \mathbb{R}^d$ and a key vector $k_m, k_n \in \mathbb{R}^d$. Next, multiply $q_m$ by the rotation matrix $\mathcal{R}_m \in \mathbb{R}^{N\times d}$, and $k_n$ by the rotation matrix $\mathcal{R}_n \in \mathbb{R}^{N\times d}$. Without loss of generality, we elaborate using a vector $q_m$ of position $m$. The above process can be described as:

$$\mathcal{R}_m \times q_m = \begin{pmatrix} \cos m\theta_0 & -\sin m\theta_0 & 0 & 0 & \cdots & 0 & 0 \\ \sin m\theta_0 & \cos m\theta_0 & 0 & 0 & \cdots & 0 & 0 \\ 0 & 0 & \cos m\theta_1 & -\sin m\theta_1 & \cdots & 0 & 0 \\ 0 & 0 & \sin m\theta_1 & \cos m\theta_1 & \cdots & 0 & 0 \\ \vdots & \vdots & \vdots & \vdots & \ddots & \vdots & \vdots \\ 0 & 0 & 0 & 0 & \cdots & \cos m\theta_{d/2-1} & -\sin m\theta_{d/2-1} \\ 0 & 0 & 0 & 0 & \cdots & \sin m\theta_{d/2-1} & \cos m\theta_{d/2-1} \end{pmatrix} \times \begin{pmatrix} q_0 \\ q_1 \\ q_2 \\ q_3 \\ \vdots \\ q_{d-2} \\ q_{d-1} \end{pmatrix} \quad (10)$$

$$\theta_i = 10000^{-2i/d}$$

During the inner product operation in relative position aware attention, it is easy to know that the following equation holds constant due to the rotation matrix property:

$$(\mathcal{R}_m q_m)^{\mathrm{T}} (\mathcal{R}_n k_n) = q_m^{\mathrm{T}} \mathcal{R}_m^{\mathrm{T}} \mathcal{R}_n k_n = q_m^{\mathrm{T}} \mathcal{R}_{n-m} k_n \quad (11)$$

Therefore, the introduction of the matrix $\mathcal{R}_{n-m}$ enables the encoding of relative temporal information when embedding features. As a result, relative time-series information on the degradation process can thus be fully utilized, contributing to long-term extrapolation prediction.

Based on $\mathcal{R}_{n-m}$, the relative temporal attention vector $e_r^{(i)}$ is computed and expressed as follows.

$$e_r^{(i)} = R_{attention}(x_p^{(i)}) = \mathrm{Softmax}\left(\frac{Q\mathcal{R}K^{\mathrm{T}}}{\sqrt{d}}\right)V$$
$$Q = \left(x_p^{(i)}\right)^{\mathrm{T}} W_{rote}^{Q}; \quad K = \left(x_p^{(i)}\right)^{\mathrm{T}} W_{rote}^{K}; \quad V = \left(x_p^{(i)}\right)^{\mathrm{T}} W_{rote}^{V} \quad (12)$$

In summary, the triple embeddings can be summed up and the resulting embedding $e_{emb}^{(i)} \in \mathbb{R}^{N\times d}$ is fed into the FPLM module, denoted as:

$$e_{emb}^{(i)} = e_{token}^{(i)} \oplus e_{pos(n,j)}^{(i)} \oplus e_r^{(i)} \quad (13)$$

### 3.3.3 FPLM Module

FPLM module takes the out of Input embedding module, $e_d^{(i)}$, and fuses the generalized token prediction knowledge from the pre-training for deep representation learning. Specifically, $e_d^{(i)}$ will be fed into the multiple stacked FPLM blocks. Then each head $h = 1, 2, ..., H$ in each block $b = 1, 2, ..., B$ of the whole FPLM module will transform them into Query Matrices $Q_{b,h}^{(i)} = \left(e_d^{(i)}\right)^T W_{b,h}^Q$, Key Matrices $K_{b,h}^{(i)} = \left(e_d^{(i)}\right)^T W_{b,h}^K$, Value Matrices $V_{b,h}^{(i)} = \left(e_d^{(i)}\right)^T W_{b,h}^V$, where $W_{b,h}^Q, W_{b,h}^K, W_{b,h}^V \in \mathbb{R}^{N \times d}$ are fixed parameter matrices, since we freeze the multi-head attention layer in each FPLM block. After that, the output $O_{b,h}^{(i)} \in \mathbb{R}^{N \times D}$ that is deeply and adaptively represented by the FPLM block incorporated with intrinsic degradation information can be written as:

$$\left(O_{b,h}^{(i)}\right)^T = \text{FPLM}_{attention}\left(Q_{l,h}^{(i)}, K_{l,h}^{(i)}, V_{l,h}^{(i)}\right) = \text{Softmax}\left(\frac{Q_{l,h}^{(i)} K_{l,h}^{(i)T}}{\sqrt{d}}\right) V_{l,h}^{(i)} \tag{14}$$

As mentioned earlier, we used a pre-trained LLM as the backbone network of FPLM-HEL. However, fully fine-tuning all parameters of the LLM would result in an excessively high computational overhead that is unbearable. For example, the GPT2 model itself has a 12-layer decoder structure, and each decoder layer is embedded with a 12-head multi-head attention layer. Excluding the frozen parameters, the parameters required for fine-tuning are still large. Therefore, to further alleviate the computational burden of fine-tuning a large number of parameters and to avoid overfitting such a deep model on degradation data, we used the first 6 layers of GPT2, consistent with the literature (Zhou et al., 2023). This paper also verifies in subsequent experiments why the choice of the first 6 layers is balanced. As shown in Figure 2, each FPLM block also has residual links and frozen feed-forward network layers, therefore, the obtained deep representations $O_{b,h}^{(i)}$ are subsequently fed into these layers in sequence to complete multiple nonlinear mappings. This process can be expressed as follows:

$$\begin{aligned}
\underbrace{H_{b,h}^{(i)}}_{start} &= \text{LayerNorm}\left(O_{b,h}^{(i)}\right) \\
\underbrace{H_{(b,h)}^{(i)}}_{end} &= \text{LayerNorm}\left\{\delta\left[\text{FeedFoward}\left(\underbrace{H_{b,h}^{(i)}}_{start}\right) + \underbrace{H_{b,h}^{(i)}}_{start}\right]\right\}
\end{aligned} \tag{15}$$

where $\underbrace{H_{b,h}^{(i)}}_{start} \in \mathbb{R}^{N \times d}$ represents the hidden layer representation after normalization, which is then input into the subsequent network layers. $\underbrace{H_{(b,h)}^{(i)}}_{end} \in \mathbb{R}^{N \times d}$ represents the final output of FPLM block, $\delta(\cdot)$ is the modified GELU (Hendrycks and Gimpel, 2023) activation function used to complete further nonlinear mapping within the feed-forward network of each FPLM. Subsequently, after stacking a total of $b$ FPLM blocks, the input degradation data is adaptively learned layer by layer and outputs high-dimensional embeddings. These embeddings are reorganized and generated by GPT2 using the general knowledge obtained during pre-training, encompassing deep degradation characteristics of the bearings. Typically, these high-order representations span a larger feature space, allowing the final Fully Connected (FC)

Layer to complete the regression task from a wide reserve of feature maps, thus effectively converging quickly in the output space. Therefore, in this study, a FC layer is used instead of more complex convolutional layers or LSTM units to complete the final mapping and estimation of RUL. The entire framework ultimately outputs the RUL values for a future period.

### 3.3.4 Two-Stage Finetuning

To effectively utilize the advantages of LLM, we use a two-stage fine-tuning strategy to gradually activate its domain adaptation capability and transfer the generalized prediction knowledge from the pre-training stage. The training process consists of two steps: first, the Supervised Fine-Tuning (SFT) stage retains all pre-trained prediction knowledge of the LLM, transferring it to the downstream RUL prediction task and fully exploiting degradation features in the source domain. Second, the Prompt Tuning (PT) stage uses a small number of degraded samples in the target domain to prompt and guide the LLM to make predictions, activating its generalization ability. As shown in Figure 4, the network layers that need to be fine-tuned and frozen at each stage are marked with different colors. To facilitate representation, the LSPR component is denoted as the operator $\mathcal{G}(\cdot)$, and the FPLM-HEL component as the RUL predictor is denoted as $\mathcal{F}(\cdot)$, the set of trainable parameters of the proposed model is $\mathcal{P}_\Theta = \{\Theta_{freeze}, \Theta_{tuning}\}$, where $\Theta_{freeze}$ denotes the frozen set of parameters, $\Theta_{tuning}$ denotes the updated set of parameters, and the parameters $\Theta_{tuning} = \{\theta_1, \theta_2\}$ that need to be updated in each of the two phases are denoted as $\theta_1$ and $\theta_2$. The detailed procedure of the two-phase fine-tuning is as follows:

1) Supervised Fine-Tuning (SFT)

In the SFT stage, using the SFT dataset $\mathcal{D}_{SFT} = \{\mathcal{X}_{SFT}, \mathcal{Y}_{SFT}\}$ collected on the source domain $\mathcal{S}$, where $\mathcal{X}_{SFT} = \{x_\mathcal{S}^{(i)}\}_{i=1}^{N_s}$ denotes the set of vibration signal samples, $\mathcal{Y}_{SFT} = \{r_\mathcal{S}^{(i)}\}_{i=1}^{N_s}$ denotes the corresponding RUL label, the loss function of this stage is shown below:

$$\mathcal{L}_{SFT} = \frac{1}{N_s} \sum_{i=1}^{N_s} (r_\mathcal{S}^{(i)} - \tilde{r}_\mathcal{S}^{(i)})^2$$
$$\tilde{r}_\mathcal{S}^{(i)} = \mathcal{F}_{\hat{\theta}_1}(\mathcal{G}(x_\mathcal{S}^{(i)}))$$
(16)

where $N_s$ is the size of the sample set in the source domain, $\hat{\theta}_1$ denotes all model parameters updated in the SFT stage.

2) Prompt Tuning (PT)

In the PT stage, using the PT dataset $\mathcal{D}_{PT} = \{\mathcal{X}_{PT}, \mathcal{Y}_{PT}\}$ collected on the target domain $\mathcal{T}$, where $\mathcal{X}_{PT} = \{x_\mathcal{T}^{(i)}\}_{i=1}^{N_t}$ denotes the set of vibration signal samples, $\mathcal{Y}_{PT} = \{r_\mathcal{T}^{(i)}\}_{i=1}^{N_t}$ denotes the corresponding RUL label, the loss function of this stage is shown as:

$$\mathcal{L}_{PT} = \frac{1}{N_t} \sum_{i=1}^{N_t} (r_T^{(i)} - \tilde{r}_T^{(i)})^2$$
$$\tilde{r}_T^{(i)} = \mathcal{F}_{\hat{\theta}_2|\hat{\theta}_1}(x_T^{(i)}) \tag{17}$$

where $N_t$ is the size of the sample set in the source domain, usually $N_s > N_t$, $\hat{\theta}_2|\hat{\theta}_1$ denotes all model parameters updated at the PT stage after SFT.

The Pseudo code for the two-stage fine-tuning is shown in Table 1.

Table 1 Algorithm Pseudo Code: Two Stage Fine-tuning for Embedding Learning

| Algorithm : Two Stage Fine-tuning for Embedding Learning |
| --- |
| **Input:** $\mathcal{P}_\Theta$, $\mathcal{D}_{SFT}$, $\mathcal{D}_{PT}$, $\mathcal{F}(\cdot)$, $\mathcal{G}(\cdot)$ |
| **Output:** $\hat{\theta}_1$, $\hat{\theta}_2$ |
| **Data Preprocessing:** $\mathcal{G}(x_S^{(i)})$, $\mathcal{G}(x_T^{(i)})$ |
| **Initialize** $\theta_1$, $\theta_2$ Batch size = $N_b$, Hyperparameter Setting |
| **For** epoch=1, 2, …, $epoch_{SFT}$, freezing $\Theta_{freeze}$ |
|     Feed $\mathcal{G}(x_S^{(i)})$ to FPLM-HEL components $\mathcal{F}_{\theta_1}(\cdot)$ |
|     Calculate RUL prediction $\tilde{r}_S^{(i)} = \mathcal{F}_{\hat{\theta}_1}(\mathcal{G}(x_S^{(i)}))$ |
|     Calculate Loss: $\mathcal{L}_{SFT} = \frac{1}{N_s}\sum_{i=1}^{N_s}(r_S^{(i)} - \tilde{r}_S^{(i)})^2$, update $\hat{\theta}_1 \leftarrow \theta_1$ until minimize $\mathcal{L}_{SFT}$ |
| **End** |
| **For** epoch = 1, 2, …, $epoch_{PT}$ |
|     Feed $\mathcal{G}(x_T^{(i)})$ to FPLM-HEL components $\mathcal{F}_{\theta_2|\hat{\theta}_1}(\cdot)$ |
|     Calculate RUL prediction $\tilde{r}_T^{(i)} = \mathcal{F}_{\hat{\theta}_2|\hat{\theta}_1}(\mathcal{G}(x_T^{(i)}))$ |
|     Calculate Loss: $\mathcal{L}_{PT} = \frac{1}{N_t}\sum_{i=1}^{N_t}(r_T^{(i)} - \tilde{r}_T^{(i)})^2$, update $\hat{\theta}_2 \leftarrow \theta_2$ until minimize $\mathcal{L}_{PT}$ |
| **End** |
| Obtain $\hat{\Theta}_{tuning} = \{\hat{\theta}_1, \hat{\theta}_2\}$, until convergence of whole model |

## 4. Experiments

In this section, we conducted two case studies, which are: 1). Comparative experiment on long term RUL transfer prediction using the FEMTO bearing dataset (Nectoux et al., 2012); 2). Analytical experiment using the XJTU-SU bearing dataset (Lei et al., 2019). In case study 1, the main purpose is to compare our proposed framework with other state-of-the-art bearing RUL prediction methods to verify its effectiveness and superiority; In case study 2, comparison experiments, ablation experiments, and analytical experiments were set up, mainly to explore the robustness of the proposed method and the impact of its key components on RUL prediction results.

## 4.1 Case Study 1: Comparative Experiments Based on The FEMTO Bearing Dataset

### 4.1.1 Data Description

The PRONOSTIA experimental platform from which the FEMTO dataset is collected is shown in the Figure 5, the PRONOSTIA experimental platform mainly consists of three parts: 1. rotating part, includes the asynchronous motor with a gearbox and its two shafts, used for transmitting the rotating motion; 2. degradation generation part, primarily includes a loading system, which induces degradation by applying force radially on the testing bearings; 3. measurement part, mainly involves the use of two radial acceleration sensors placed at 90 degrees to each other on the outer ring of the bearing to collect vibration signals. To avoid bearing failure affecting and damaging the entire test bench, the experiment will stop when the signal amplitude collected by the vibration sensor exceeds 20g. The dataset covers the natural degradation process of the bearings under three working conditions, and no artificial faults have been added, thus the FEMTO dataset fully simulates the RUL prediction requirements in real industrial scenarios. For a total of 17 bearings, the duration of the experiment ranges from 1h to 7h, making accurate RUL estimation a challenging task. The relevant algorithm runs in 11th Gen Intel(R) Core(TM) i7–11800H @ 2.30 GHz, RAM 64.0 GB, RTX4090Ti×2, and PyTorch 1.9.0.

In the FEMTO dataset, the number of bearings under the three conditions are: 7, 7, and 3, respectively, the bearing codes under condition 1 are 1-1 to 1-7, under condition 2 are 2-1 to 2-7, and under condition 3 are 3-1 to 3-3. Table 2 lists these details.

Table 2 Details of the FEMTO dataset

| Each sampling duration (s) | 0.1 | Sampling interval (s) | 10 | Sampling frequency (kHz) | 25.6 |
|---|---|---|---|---|---|
| | | **Working Condition** | | **Bearing ID** | |
| | | Load (KN) | Speed (rpm) | | |
| **FEMTO Dataset** | Condition 1 | 4.0 | 1800 | Bearing 1-1; Bearing 1-2<br>Bearing 1-3; Bearing 1-4<br>Bearing 1-5; Bearing 1-6<br>Bearing 1-7 | |
| | Condition 2 | 4.2 | 1500 | Bearing 2-1; Bearing 2-2<br>Bearing 2-3; Bearing 2-4<br>Bearing 2-5; Bearing 2-6<br>Bearing 2-7 | |
| | Condition 3 | 5.0 | 1500 | Bearing 3-1; Bearing 3-2<br>Bearing 3-3 | |

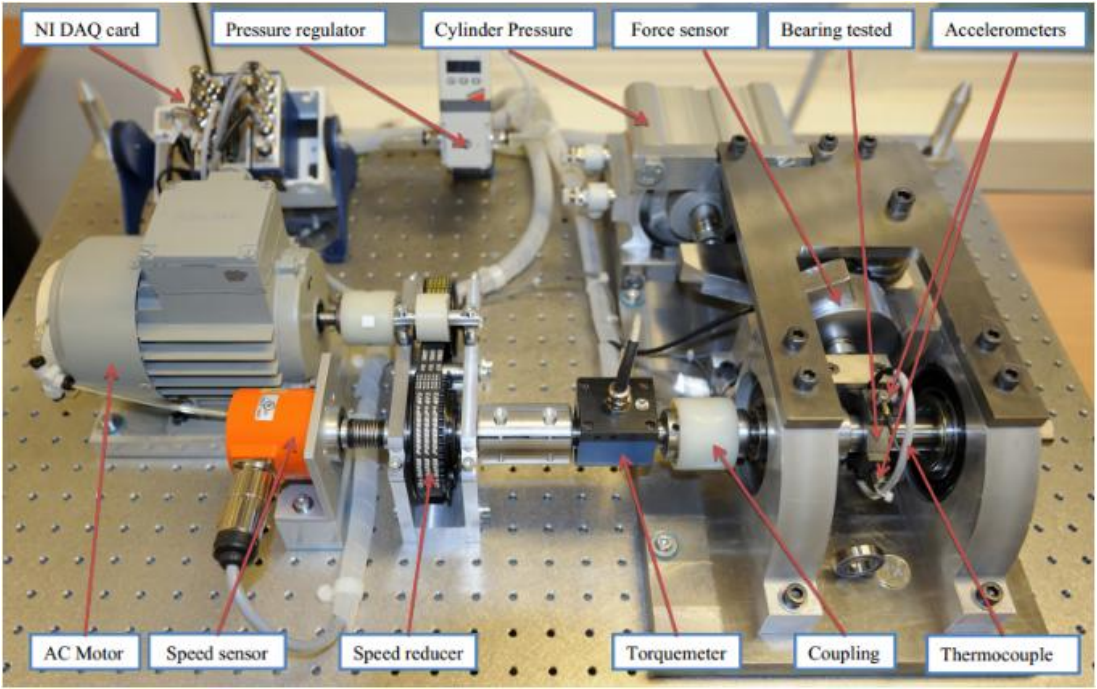

Figure 5 Overview of the PRONOSTIA experiment platform

### 4.1.2 Evaluation Metrics

To compare the RUL prediction performance of different methods, this paper uses three classical regression metrics for quantitative evaluation: MAE, RMSE, and MAPE. The specific introductions are as follows:

MAE：mean absolute error, formulated as:

$$MAE = \frac{1}{N_x} \sum_{t=1}^{N_x} \left| \widehat{RUL_t} - RUL_t \right| \tag{18}$$

RMSE：root mean square error, formulated as:

$$RMSE = \sqrt{\frac{1}{N_s} \sum_{t=1}^{N_s} \left( \widehat{RUL_t} - RUL_t \right)^2} \tag{19}$$

MAPE：mean absolute percentage error, formulated as:

$$MAPE = \frac{100\%}{N_s} \sum_{t=1}^{N_s} \frac{\left| \widehat{RUL_t} - RUL_t \right|}{\left| \widehat{RUL_t} \right|} \tag{20}$$

where represents the number of samples during testing, $\widehat{RUL_t}$ and $RUL_t$ represent the predicted RUL value and the actual RUL value at the $t$-th prediction time step, respectively. MAE provides an intuitive understanding of the absolute value of prediction errors, while RMSE takes the square root of MSE, aligning the error unit with the original data unit and giving higher weight to larger errors. MAPE is a percentage-based metric, making it suitable for comparing the prediction performance of datasets of different scales. Clearly, for these three quantitative evaluation metrics, smaller values are better.

### 4.1.3 Configuration of The Experiments

(1) Data preprocess

During the data preprocessing stage, we use the Short-Time Fourier Transform (STFT) method to extract time-frequency domain features from the lateral vibration signals (Oberlin et al., 2014). Specifically, we set the length of the sliding window to 2560, indicating that the size of each original sequence is 1×2560. In STFT, the frame width and frame stride are set to 20ms and 10ms respectively, with these empirical hyperparameter choices coming from the literature(Su et al., 2021). Thus, each original sequence, after STFT transformation, yields a feature matrix of size 11×256. Furthermore, we compress this feature matrix to a feature vector of dimension 64. In other words, each segment of the original vibration signal of length 2560, after preprocessing, becomes a feature vector of size 1×64, containing the time-frequency domain information of the bearing degradation process. These extracted time-frequency domain features are then stacked together in the sliding window direction to provide the basic data input for the proposed RUL prediction framework. Since the lifespan range of bearings in the experiments is broad, using the actual RUL as labels would result in large gradient changes in the network, potentially leading to model underfitting. Therefore, we normalize the RUL values of different sampling points by dividing them by their respective total lifespan, scaling the large range of RUL values to the [0, 1] interval. In other words, the percentage of RUL is used as the target output of the proposed framework.

(2) Hyperparameters setting

In the proposed RUL prediction framework, the hyperparameters that need to be predefined include the patch size, the stride between patches, and the prediction length, etc. Table 3 lists the specific choices of hyperparameters. In particular, in Section 4.2.3, we discuss in detail the impact of some hyperparameter choices, including the trade-off between patch size and patch stride, as well as the impact of prediction length on the prediction results.

Table 3 The hyperparameter settings

| Hyperparameter | Value | Hyperparameter | Value |
|---|---|---|---|
| Optimizer | Adam | Historical step length | 75 |
| Learning rate | 1e-5 | Batch size | 50 |
| Loss function | MSE | Patch size | 6 |
| Dropout rate | 0.2 | Patch stride | 4 |
| Prediction step length | 25 | Block number $b$ | 6 |
| Hidden size | 768 | SFT epoch | 64 |
| PT epoch | 16 | | |

(3) Comparison SOTA methods

To validate the effectiveness and superiority of the proposed method, we compared our model with four widely applied typical methods related to shallow machine learning, deep learning, and transfer learning, including Support Vector Regression (SVR) (Zhang & O'Donnell, 2020), Long Short-Term Memory Networks (LSTM) (Hochreiter & Schmidhuber, 1997), Temporal Convolutional Networks (TCN) (Bai et al., 2018), and Deep Domain Adaptive Network (DDAN) (Miao & Yu, 2021). For fairness, hyperparameter settings related to the training process, such as the number of epochs and batch size, are the same for all methods. Additionally, the architectural settings of deep models can show significant differences in results. Therefore, for these methods, we referred to the specific values used in the relevant papers to ensure the fairness of the comparative experimental results. For example, we chose the classic three-layer architecture for TCN and LSTM; for DDAN, we adhered to the parameter values set in the original literature and used them in our experiments.

For each method, we designed 6 transfer RUL prediction tasks representing the transfer from source domain data of different distributions to the target domain. Each method conducts the corresponding RUL prediction experiments on these 6 types of tasks. The detailed information on the 6 transfer prediction tasks is shown in Table 4.

Table 4 Experimental task setup, including details of source domain and target domain data

| Task | Transfer path | | Training phase | | Testing phase |
|---|---|---|---|---|---|
| | Source condition | Target condition | SFT set | Prompting set | Test set |
| 1 | C1 | C3 | Bearing 1-1<br>Bearing 1-2 | Bearing 3-2 | Bearing 3-3 |
| 2 | C2 | C3 | Bearing 2-1<br>Bearing 2-2 | Bearing 3-2 | Bearing 3-3 |
| 3 | C3 | C1 | Bearing 3-1<br>Bearing 3-2 | Bearing 1-1 | Bearing 1-3 |
| 4 | C2 | C1 | Bearing 2-1<br>Bearing 2-2 | Bearing 1-1 | Bearing 1-3 |
| 5 | C3 | C2 | Bearing 3-1<br>Bearing 3-2 | Bearing 2-1 | Bearing 2-3 |
| 6 | C1 | C2 | Bearing 1-1<br>Bearing 1-2 | Bearing 2-1 | Bearing 2-3 |

### 4.1.4 Experiments Results and Discussion

In this case study, we first compared the proposed bearing RUL prediction framework with four representative methods under the set transfer tasks to demonstrate its superiority. Then, by visualizing the RUL prediction curves, we deeply analyzed and discussed the performance of different models and the advantages of the proposed method under stringent long-term prediction conditions. Specifically, in this case study, each task was applied 10-folds cross validation.

As shown in Table 5, values in bold black represent the best prediction results among all methods, while the second-best results are underlined. It is clear that, under the three evaluation metrics, the proposed RUL prediction framework outperforms other comparative methods in all six cross-condition RUL prediction tasks. Compared to the domain adaptation method DANN, the proposed framework's performance improves by at least 16.22% (in Task 3 under the MAE metric). In the six experimental tasks set, the methods using transfer learning significantly outperform the conventional data-driven methods without transfer learning, confirming the necessity of using transfer learning methods under conditions of differentiated data distribution.

Additionally, as representatives of deep neural networks, LSTM and TCN outperform SVR in prediction results. This can be explained by the fact that deep neural networks can better extract useful degradation features from training data compared to shallow machine learning methods. LSTM and TCN, through their hierarchical network structures, can capture complex patterns in time series data, particularly excelling in handling long-term dependencies and nonlinear dynamic changes. As a traditional machine learning method, SVR has certain advantages in handling linear regression problems, but it has limitations in capturing complex nonlinear relationships and long-term dependencies. o more intuitively compare the experimental results of different methods, Figure 6 demonstrates the superiority of the proposed method. These charts can more clearly convey the performance differences of various models in long-term RUL prediction tasks.

To further analyze the performance of different methods under long-term prediction settings, Figure 7 shows the specific prediction results at different time steps, we illustrate this by taking Bearing 3-3 in Task 6 as example. It can be clearly seen that in the final segment of RUL prediction output, all prediction methods show an upward trend, indicating that long-term prediction remains a significant challenge. The reason behind this phenomenon is that the degradation behavior in the later stages of bearing life is more severe and complex, with increased noise and uncertainty in the signals, making model prediction more difficult. Notably, compared to other methods, the proposed method is as close as possible to the actual RUL value with the smallest deviation, followed by DDAN, while the other three prediction methods show a trend of significantly deviating from the actual RUL value in later prediction outputs. This phenomenon further indicates that the proposed method has significant advantages in handling long-term dependencies and complex degradation patterns.

In-depth, the reason the proposed method can suppress prediction differences lies in its larger model capacity and deep feature representation capability. By learning rich features during the pre-training stage, the model can capture key information in the degradation process and further optimize prediction performance through targeted adjustments during the fine-tuning stage. Our approach not only enhances the model's robustness but also strengthens its stability in long-term prediction tasks. Results indicate that in practical engineering, proposed framework has broad application prospects in RUL prediction, especially under conditions where there are significant differences in cross-condition data distribution and the need for accurate long-term predictions.

Table 5 comparison of quantitative metric results between the proposed framework and SVR, LSTM, TCN, and DDAN

| Metrics | Task | Method | | | | |
|---|---|---|---|---|---|---|
| | | Proposed framework | SVR | LSTM | TCN | DANN |
| RMSE | 1 | **0.1172** | 0.672 | 0.597 | 0.320 | 0.274 |
| | 2 | **0.0923** | 0.593 | 0.377 | 0.297 | 0.228 |
| | 3 | **0.0877** | 0.516 | 0.349 | 0.324 | 0.150 |
| | 4 | **0.0782** | 0.609 | 0.310 | 0.291 | 0.113 |
| | 5 | **0.0952** | 0.799 | 0.350 | 0.347 | 0.165 |
| | 6 | **0.0873** | 0.713 | 0.472 | 0.361 | 0.216 |
| MAE | 1 | **0.1043** | 0.670 | 0.510 | 0.317 | 0.241 |
| | 2 | **0.0901** | 0.589 | 0.472 | 0.241 | 0.197 |
| | 3 | **0.0821** | 0.509 | 0.430 | 0.298 | 0.098 |
| | 4 | **0.0721** | 0.597 | 0.393 | 0.252 | 0.109 |
| | 5 | **0.0896** | 0.791 | 0.477 | 0.322 | 0.157 |
| | 6 | **0.0835** | 0.704 | 0.491 | 0.349 | 0.207 |
| MAPE | 1 | **0.3621** | 1.651 | 0.810 | 0.550 | 0.471 |
| | 2 | **0.2913** | 0.975 | 0.739 | 0.510 | 0.442 |
| | 3 | **0.2276** | 0.901 | 0.800 | 0.448 | 0.350 |
| | 4 | **0.2113** | 1.032 | 0.695 | 0.371 | 0.290 |
| | 5 | **0.2998** | 1.773 | 0.845 | 0.460 | 0.393 |
| | 6 | **0.3356** | 1.720 | 0.868 | 0.509 | 0.410 |

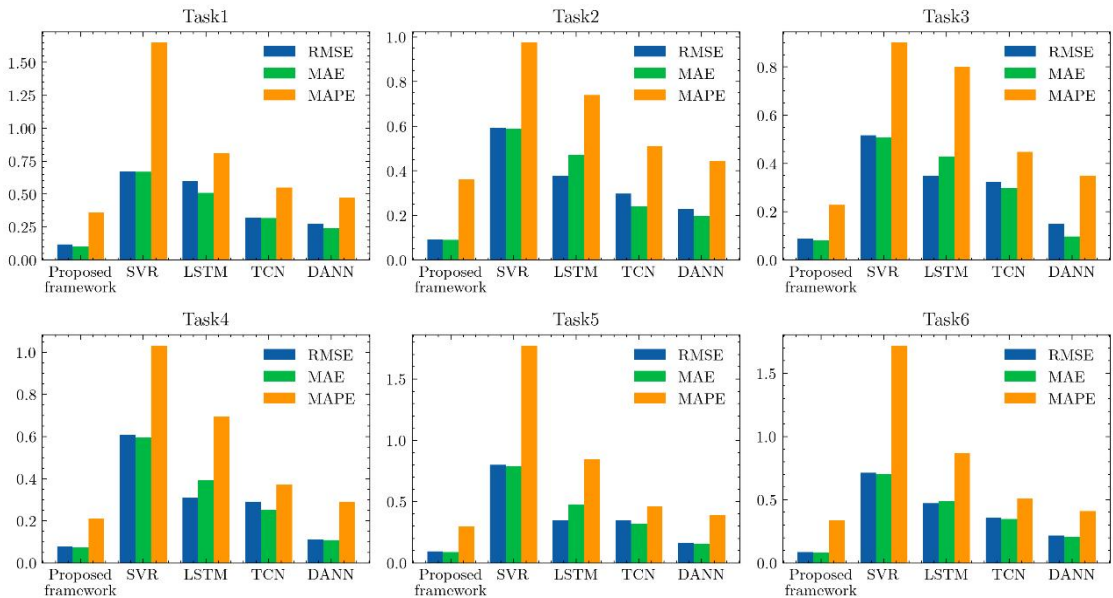

Figure 6 Comparison of the proposed framework with SVR, LSTM, TCN, and DDAN on RUL prediction results using three evaluation metrics

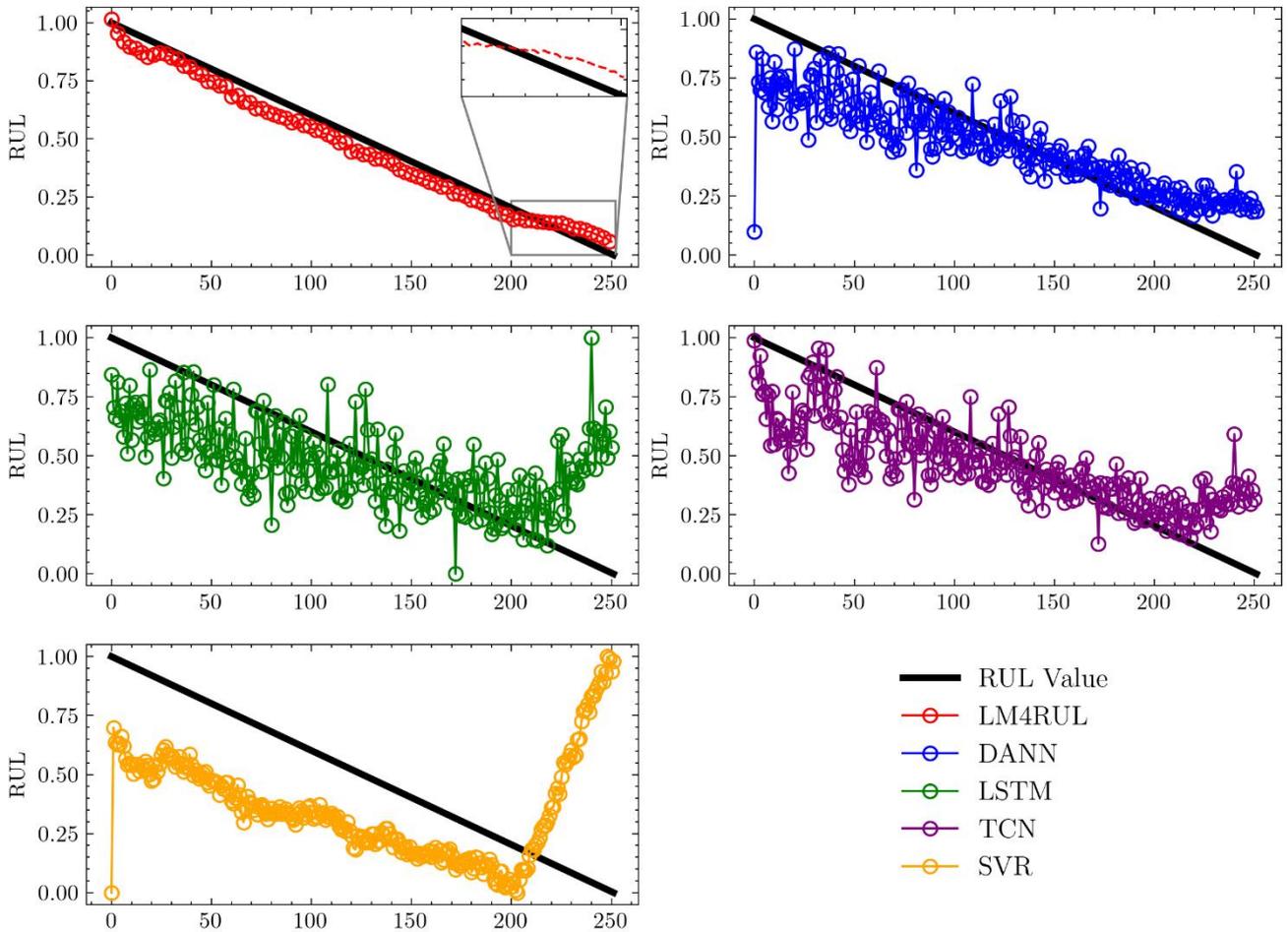

Figure 7 RUL prediction results of five comparative methods for bearing 3-3 in the FEMTO dataset

## 4.2 Case Study 2：Towards the understanding of PLM's general prediction capability based on XJTU-SY

### 4.2.1　Data Description and Evaluation Metrics

For the core components and important hyperparameters included in our proposed framework, to further discuss their impact on RUL prediction, this section uses another well-known accelerated degradation test dataset for bearing life: XJTU-SY. Because the specific failure behavior of each bearing in the XJTU-SY dataset is explicitly provided compared to the FEMTO dataset, this helps to conduct more in-depth analytical experiments. The XJTU-SY rolling bearing accelerated life testing platform consists of an AC motor, motor speed controller, shaft, support bearings, hydraulic loading system, and test bearings. The adjustable conditions of the test platform mainly include radial force and rotational speed, where the radial force is generated by the hydraulic loading system acting on the bearing seat of the test bearing, and the speed is set and adjusted by the speed controller of the AC motor. To obtain the full lifecycle vibration signals of the bearing, two PCB 352C33 unidirectional accelerometers are fixed to the horizontal and vertical directions of the test bearing using magnetic bases. The DT9837 portable dynamic signal acquisition system is used to collect vibration signals during the tests. The detailed working conditions and sampling information of the XJTU-SY rolling bearing dataset can be seen in the table below. This dataset includes three operating conditions, under each of which five run-to-failure data of LDK UER204 ball bearings were collected. The bearings under the first two conditions mainly exhibit a single fault mode, while the bearings under the third condition mainly exhibit a combination of fault modes. The composition details of the test bench are shown in Figure 8, and the detailed information of the dataset is shown in Table 6.

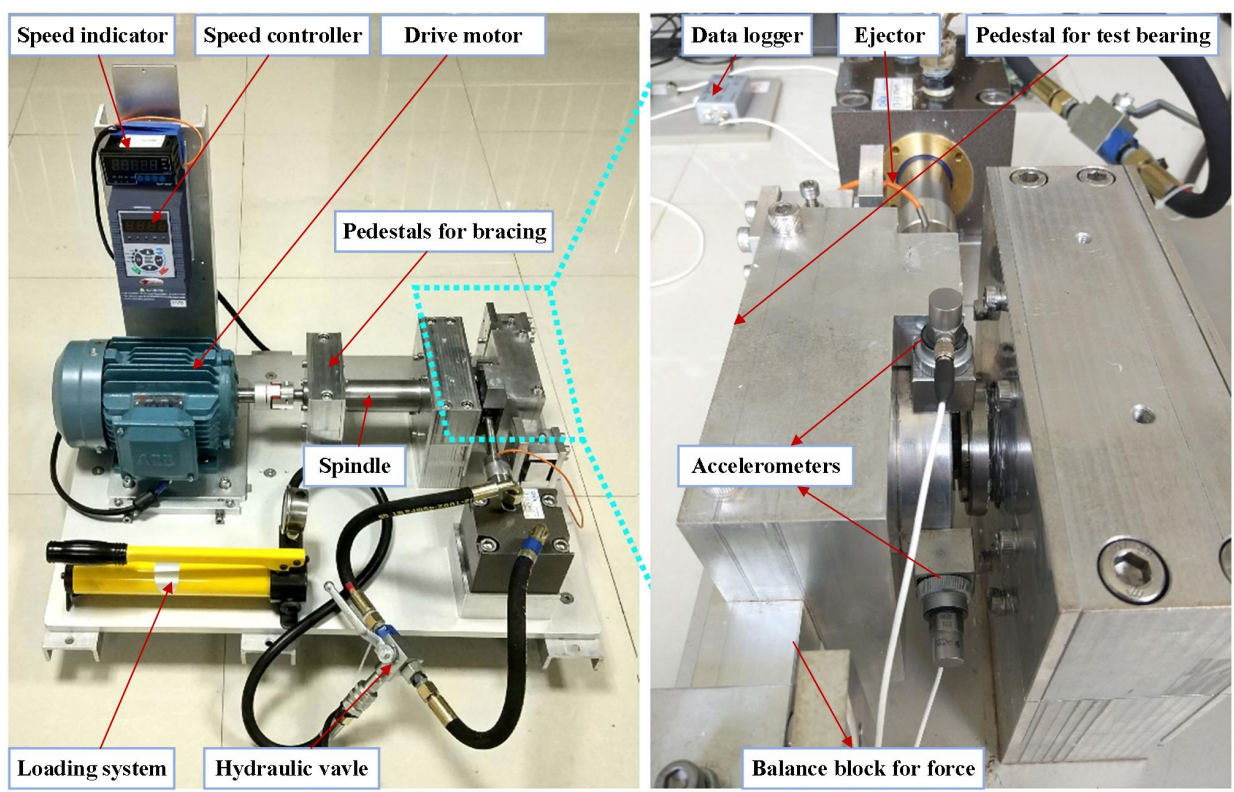

Figure 8 Detailed composition of the XJTU-SY rolling bearing accelerated degradation life test platform

Table 6 Details of the XJTU-SY datasets

| Each sampling duration (s) | 1.28 | Sampling interval (s) | 60 | Sampling frequency (kHz) | 25.6 |
|---|---|---|---|---|---|
| | | **Working Condition** | | **Bearing ID** | |
| | | Load (KN) | Speed (rpm) | | |
| **XJTU-SY Dataset** | Condition 1 | 12.0 | 1800 | Bearing 1-1; Bearing 1-2 Bearing 1-3; Bearing 1-4 Bearing 1-5 | |
| | Condition 2 | 11.0 | 1500 | Bearing 2-1; Bearing 2-2 Bearing 2-3; Bearing 2-4 | |

|               |             |      | Bearing 2-5              |
|---------------|-------------|------|--------------------------|
|               |             |      | Bearing 3-1; Bearing 3-2 |
| Condition 3   | 10.0        | 1500 | Bearing 3-3; Bearing 3-4 |
|               |             |      | Bearing 3-5              |

In this part of the experiment, we additionally used another recognized evaluation metric. This is because, in RUL prediction, later predictions are more likely to lead to severe failure accidents, so later RUL predictions should be penalized more heavily than earlier predictions. The three quantitative metrics used in Section 4.1.2 mainly evaluate the overall regression effect. To take the aforementioned penalties into account and to more comprehensively evaluate the prediction performance of LM4RUL, this part of the experiment uses the score function to evaluate the RUL prediction performance under different experimental tasks. The score function used in this paper is derived from the 2008 Prognostics and Health Management Data Challenge (Saxena and Goebel, 2008), and its calculation method is shown in the following formula:

$$score = \begin{cases} \sum_{t=1}^{N_x} \left( e^{-\frac{d_t}{13}} - 1 \right), \text{ when } d_t < 0 \\ \sum_{t=1}^{N_x} \left( e^{\frac{d_t}{10}} - 1 \right), \text{ when } d_t \geq 0 \end{cases} \quad (21)$$

$$d_t = \widehat{RUL_t} - RUL_t$$

In which $d_t$ represents the difference between the predicted RUL value and the true RUL value. Taking the range of -50 to 50 as an example, the scoring function is shown in Figure 9. Unlike the RMSE evaluation metric, the scoring function overall exhibits an asymmetric and exponential growth form. When the prediction error is smaller, the score is lower, and due to the asymmetry, later predictions ($d_t < 0$) are penalized more heavily than earlier predictions. By adopting this additional evaluation metric, we aim to provide a more rigorous and comprehensive performance assessment to ensure the reliability and safety of RUL predictions in practical applications.

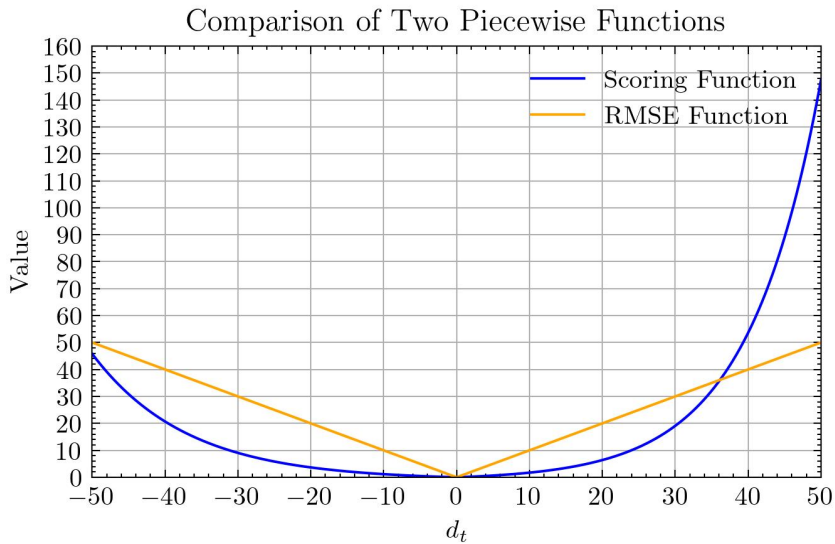

Figure 9 Comparison of the score function and RMSE metric function graphs

### 4.2.2 Comparative Experiments Results

As shown in Table 7, based on the XJTU-SY dataset, we set up three cross-condition RUL prediction tasks to comprehensively examine the long-term prediction capabilities of the proposed LM4RUL framework under these tasks. Apart from the prediction length, LM4RUL uses the same configuration as in Table 3.

Table 7 Details of the experimental task setup, including the source domain and target domain datasets

| Task | Transfer path | | Training phase | | Testing phase |
|---|---|---|---|---|---|
| | Source condition | Target condition | SFT set | Prompting set | Test set |
| 1 | C2, C3 | C1 | Bearing 2-1, Bearing 3-1<br>Bearing 2-2, Bearing 3-1<br>Bearing 2-3, Bearing 3-1 | Bearing 1-2<br>Bearing 1-3 | Bearing 1-4<br>Bearing 1-5 |
| 2 | C1, C3 | C2 | Bearing 1-1, Bearing 3-1<br>Bearing 1-2, Bearing 3-2<br>Bearing 1-3, Bearing 3-3 | Bearing 2-2<br>Bearing 2-3 | Bearing 2-4<br>Bearing 2-5 |
| 3 | C1, C2 | C3 | Bearing 1-1, Bearing 2-1<br>Bearing 1-2, Bearing 2-2<br>Bearing 1-3, Bearing 2-3 | Bearing 3-2<br>Bearing 3-3 | Bearing 3-4<br>Bearing 3-5 |

Table 8 and Figure 10 show the comparison results of the proposed LM4RUL model and four other advanced data-driven RUL prediction methods on the XJTU-SY dataset. The best prediction metric values are highlighted in bold, and the second-best are underlined. It is evident that in all experimental tasks, the proposed LM4RUL achieved the lowest score values. Compared to the DDAN method, its prediction performance improved by an average of (28.63% + 37.26% + 44.51%)/3 = 36.8% across the three prediction tasks. This indicates that the proposed LM4RUL also achieved the best prediction performance on the XJTU-SY dataset compared to the other methods. Specifically, by repeating the experiments multiple times, we aim to obtain statistically significant comparative results. As shown in Figure 10, we collected the metric results and their distribution error bands from 10 runs of each method. It can be observed that the proposed LM4RUL not only outperforms the comparative methods in cross-condition long-term RUL prediction accuracy but also maintains an advantage in prediction stability. Another observation is that all methods perform best in Task 1, followed by Task 2, and worst in Task 3. This is because Task 3 evaluates these RUL prediction methods' ability to transfer from conditions 1 and 2 to condition 3. Bearings in condition 3 mostly exhibit complex degradation behaviors under compound faults, making accurate RUL prediction more challenging. Consequently, all methods show higher score values in Task 3 compared to the other tasks.

To further analyze the predictive capabilities of the proposed LM4RUL framework, we illustrate the RUL prediction results of bearing 2-5 under Task 2 as an example. As shown in Figure 11, all methods' predicted RUL values differ significantly from the actual RUL labels during the initial degradation phase of the bearing and are almost all lower than the actual RUL labels. This is acceptable as the bearing has not yet experienced severe faults in the early stage of its lifespan. However, in the more critical later stage of its lifespan, the proposed method's prediction performance surpasses all comparative methods. Notably, in the final phase, shallow machine learning methods (SVR) and deep neural networks without transfer learning strategies (TCN, LSTM) show a distinctly opposite trend in their prediction results compared to the actual RUL labels. In the later stages of severe bearing degradation, incorrect RUL trend predictions can misguide predictive maintenance, leading to catastrophic failures and consequences. Additionally, in Figure 11, we magnified the actual late-stage prediction scenarios of the proposed method and the competitive DDAN to report their differences. It can be observed that LM4RUL's predicted RUL not only has smaller numerical deviations from the labels but also aligns more closely with the actual life trend. In Figure 12, we visualized the high-dimensional degradation features learned by LM4RUL in the source and target domains using the t-SNE algorithm for dimensionality reduction, to further investigate the domain transfer capabilities of the proposed LM4RUL. Specifically, it can be seen that under the three tasks, the proposed LM4RUL can converge in the wide feature map of the source domain and then learn the change trajectory of the bearings in the target domain. All present a continuous manifold state, indicating that LM4RUL has keenly captured the intrinsic degradation trajectories of bearings from different domains. Combining the above analysis, it is evident that this is due to the large memory capacity advantage of the large model architecture, and the two-stage fine-tuning method proposed in this paper has stepwise activated this cross-domain transfer capability.

Table 8 RUL prediction results of the proposed LM4RUL framework compared with other methods

| Metrics | Task | Method | | | | |
| --- | --- | --- | --- | --- | --- | --- |
| | | LM4RUL | SVR | LSTM | TCN | DANN |
| Score | 1 | **105.37** | 1086.70 | 976.54 | 476.82 | <u>147.64</u> |
| | 2 | **133.25** | 1632.05 | 1344.74 | 673.52 | <u>212.40</u> |
| | 3 | **451.42** | 2117.26 | 1632.48 | 971.50 | <u>813.56</u> |

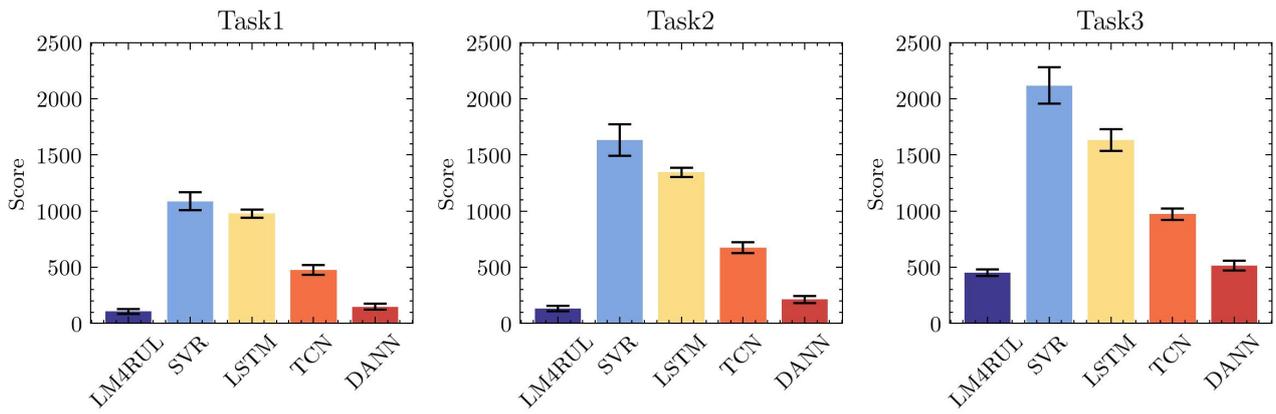

Figure 10 Score function metrics and prediction stability of RUL predictions by the proposed LM4RUL framework and

comparison methods

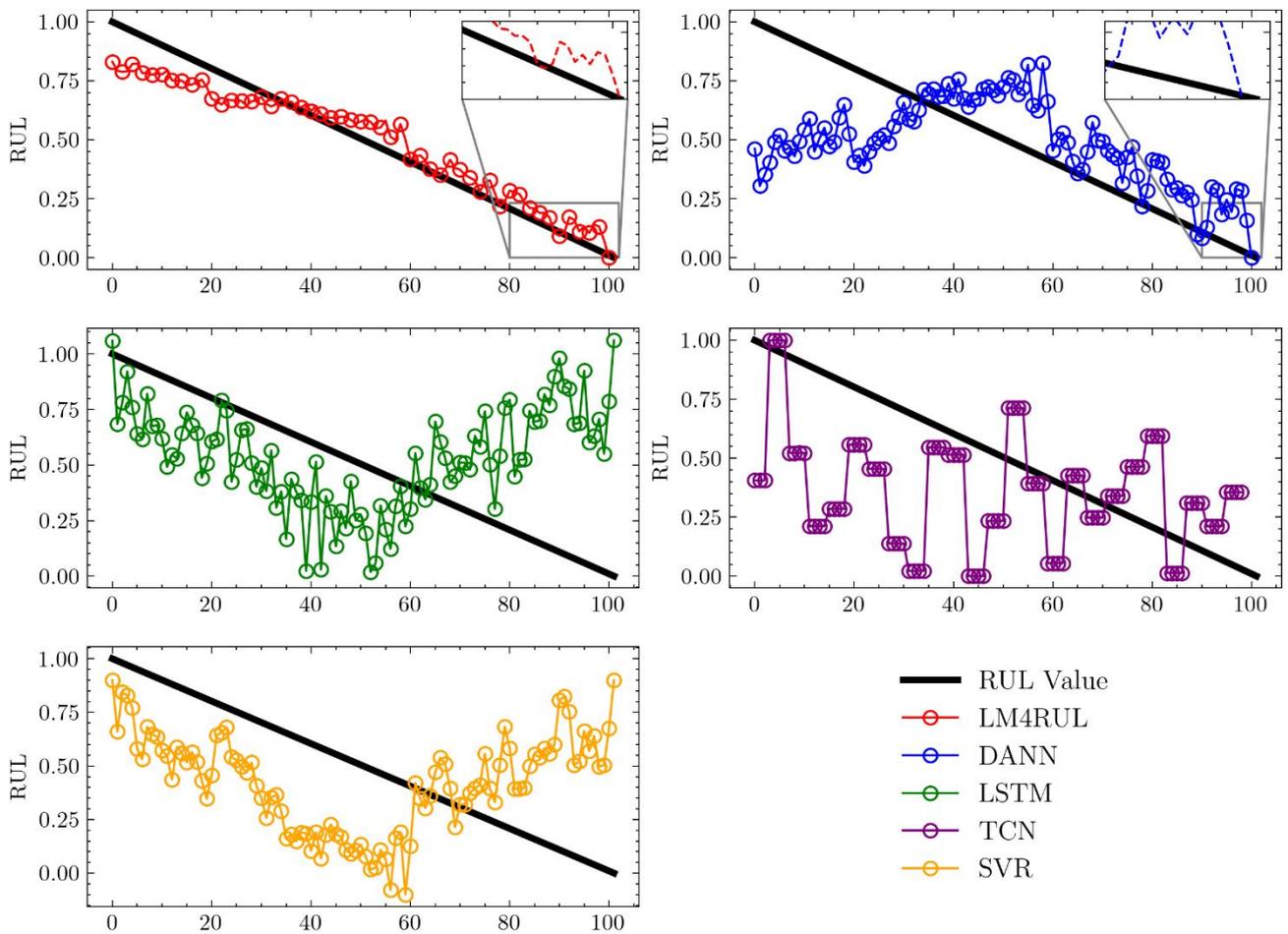

Figure 11 RUL prediction results of five comparative methods for bearing 2-5 in the XJTU-SY dataset

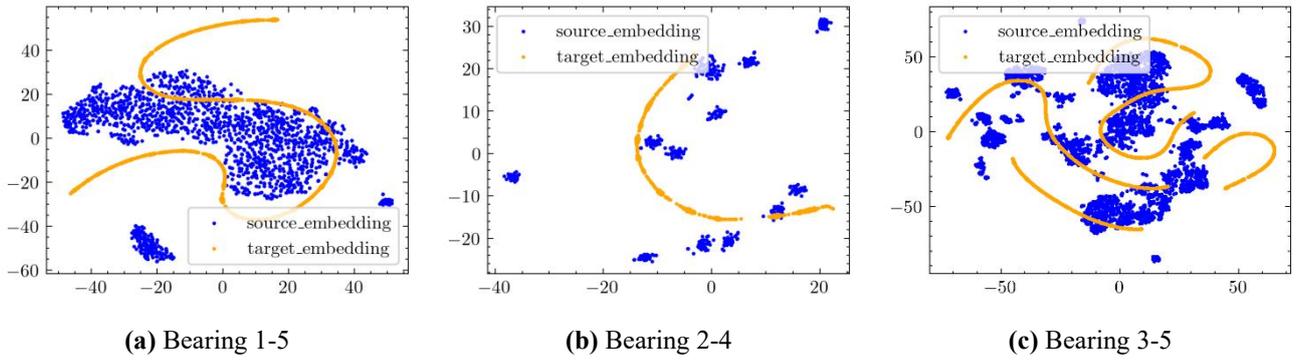

| (a) Bearing 1-5 | (b) Bearing 2-4 | (c) Bearing 3-5 |

Figure 12 t-SNE dimensionality reduction visualization of feature distributions for source and target domains in three experimental tasks

### 4.2.3 Ablation Experiments on Key Components of Proposed Framework

To discuss the impact of the core components and hyperparameters in our proposed framework on RUL prediction, we conducted ablation and analysis experiments in this section. Specifically, based on the XJTU-SY dataset, we first conducted experiments by stacking different numbers of FPLM layers as model variants, recording their score values and corresponding time consumption in prediction task 1. Secondly, we explored the impact of three key hyperparameters on RUL prediction: patch_size, patch_stride, and prediction length, where patch_size and patch_stride collaboratively influence the performance of the LSPR component.

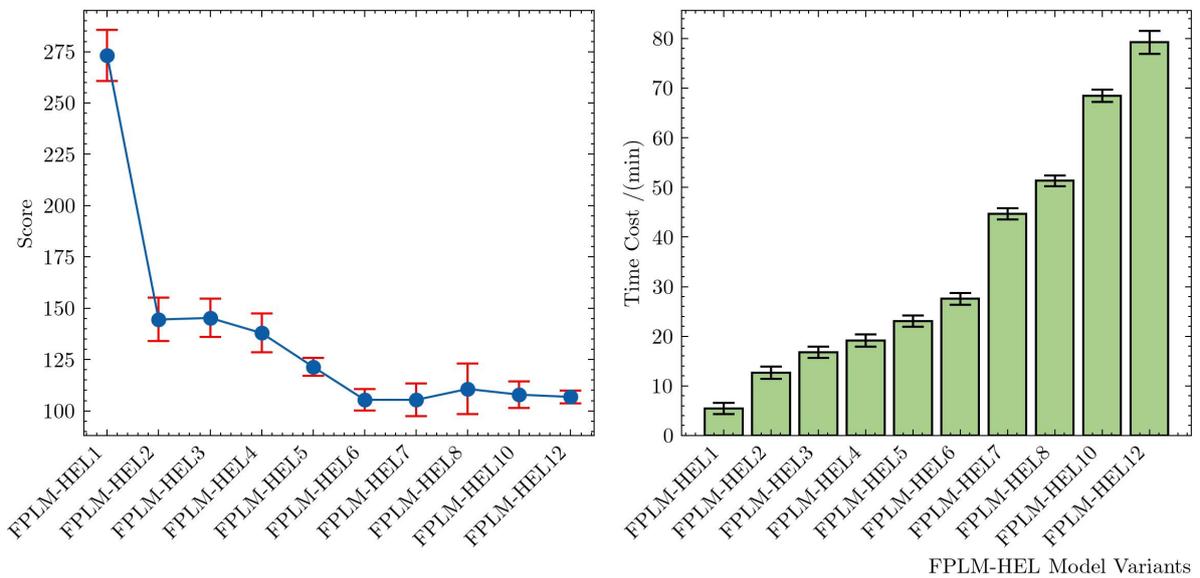

Figure 13 Comparison of prediction results and runtime of different model variants

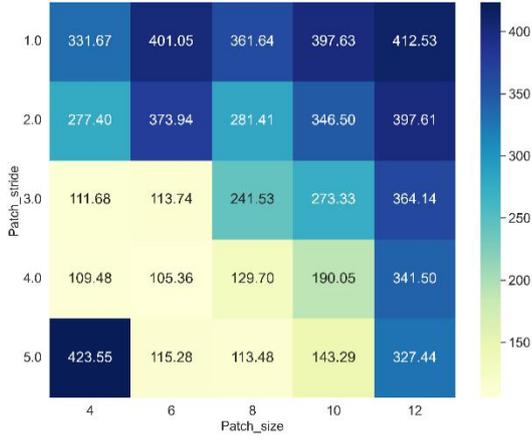 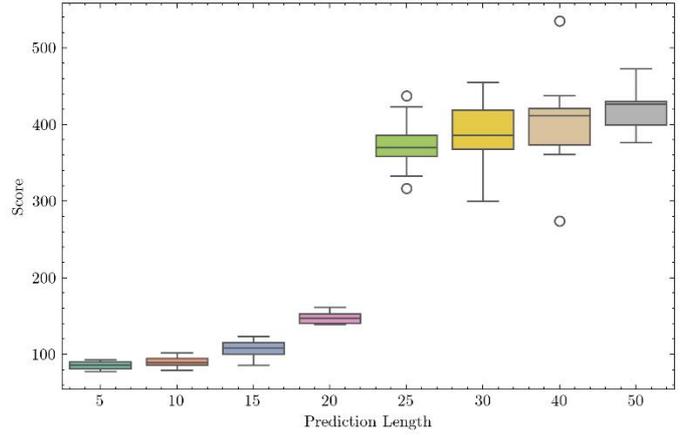

(a)            (b)

Figure 14 The impact of key hyperparameters on prediction performance: **(a)** Heatmap of the relationship between prediction results and changes in patch_stride and patch_size; **(b)** The impact of different prediction lengths on prediction results

As shown in Figure 13, it can be observed that as more FPLM-HEL layers are stacked, the score value gradually decreases. Notably, the more FPLM-HEL layers are stacked, the more pre-trained prediction knowledge is frozen. Clearly, more pre-trained prediction knowledge helps improve the RUL prediction task's effectiveness, even though the FPLM used in the pre-training stage is trained on a large amount of text. We infer this based on the conclusions of the study by Zhou et al. (Zhou et al., 2023)：large language models trained on massive textual corpora follow an autoregressive long-window token generation method during the pre-training stage, with textual corpora participating in the actual training process in the form of vectorized embeddings, which are essentially represented numerically. Therefore, during the pre-training process, large language models are likely to abstract general predictive rules independent of the text through frequent and massive fitting processes. We tend to refer to these as general predictive knowledge, formalized as parameter matrices in the attention layers of large language models. This also explains why we freeze all parameters in the attention layers and only fine-tune the parameters of other layers—general predictive knowledge should be retained as a whole when its specific mechanism of action is unclear. Combined with our experiments results, this general predictive knowledge has a high degree of extensibility and transferability, enhancing downstream RUL prediction performance. Additionally, Figure 13 reports the cost of stacking multiple FPLM-HEL layers. When six layers are stacked, the corresponding model variant achieves an optimal balance between score value and runtime. Stacking more FPLM-HEL layers does not significantly improve performance, while the runtime shows an exponential increase.

In Figure 14-(a), by experimenting with different combinations of patch_stride and patch_size values, it can be observed that the score value is minimized when the patch_stride and patch_size combination is 6, 4.

As patch_size increases and patch_stride decreases, the score value gradually increases. This experimental result can be explained by the fact that for a fixed-length historical degradation sequence, choosing a larger patch_size for patching results in a shorter historical patch sequence length. Therefore, the model needs to model key RUL information from a shorter historical patch sequence, which poses a risk of overfitting and thus leads to poor prediction performance. Meanwhile, a smaller patch_stride means that the overlapping part between patches is too large. Excessive overlapping degradation information on the patch scale also increases the difficulty of modeling RUL. Therefore, based on the above analysis, we set the patch_stride and patch_size combination to [6, 4] in the RUL prediction of bearings.

In Figure 14-(b), we studied the impact of different prediction lengths on the prediction performance of LM4RUL. It can be observed that as the desired prediction length increases, the score value continuously increases, indicating that prediction performance worsens. This is consistent with the results in the literature (Chen et al., 2023), as long-term prediction has always been recognized as a challenging problem. From these box plots, it can be seen that although shorter prediction lengths can ensure prediction accuracy, this improvement is not significant for the proposed LM4RUL. Therefore, to balance model accuracy and the time cost, this paper sets the prediction length to 20.

### 4.2.4 Understanding The Intrinsic Relationship Between The General Predictive Ability of FPLM and PCA

To gain a deeper understanding of the reasons behind the effective feature extraction behavior of FPLM-HEL, in this study, we replaced the pre-trained self-attention layers in the FPLM module with PCA and input the vibration data of bearings into the model, observing the distribution of the feature maps output by the last layer of the FPLM module on a two-dimensional plane. Taking Bearing 1-5, Bearing 2-4, and Bearing 3-5 as examples. As shown in the Figure 15, we found that the feature maps output by PCA replacing the self-attention layers have a significant consistency in distribution with the original feature maps output by FPLM. Especially for Bearing 2-5, the output feature maps highly match in trend and shape. This indicates that the self-attention layers and PCA layers probably perform similar operations, which is consistent with the findings in(Zhou et al., 2023)

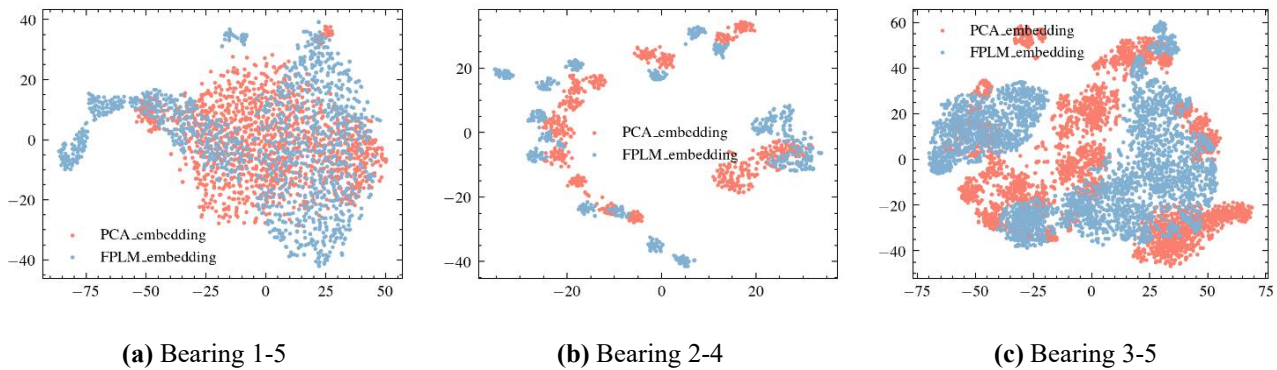

(a) Bearing 1-5  (b) Bearing 2-4  (c) Bearing 3-5

Figure 15 t-SNE dimensionality reduction comparison of bearing degradation output features between PCA and FPLM-HEL in three experimental tasks

In this study, when randomly initializing the frozen self-attention parameters with different Gaussian distribution parameters, we calculated the cosine similarity of the output feature maps from different layers of the FPLM and visualized it as a distribution histogram. To avoid excessive computational resource consumption, we selectively illustrated the cosine similarity statistics of the output feature maps from layers 1, 2, 4, and 6 in an arithmetic progression. In the graph, the horizontal axis represents the similarity interval, and the vertical axis represents the proportion of the count within a certain interval. The experimental results show that when all the self-attention layer parameters in the FPLM module are randomly initialized with Gaussian distributions, the similarity of the output feature maps on the XJTU-SY dataset is low. However, as we gradually switch to the FPLM with frozen parameters, the similarity of the output feature maps from the same layer gradually increases. Especially when all self-attention parameters are frozen, the similarity of the output feature maps from the sixth (i.e., the last) layer of the FPLM is almost entirely distributed between 90% and 100%.

This phenomenon can be explained by the fact that after LSPR, all degradation representation vectors are projected into the low-dimensional top feature vector space of the input data pattern, which is very similar to the mechanism of PCA. We further replaced the self-attention layers in the FPLM with PCA and found that not only in the sixth layer, but the similarity distribution of the output feature maps from other layers was also highly similar. This indicates that the frozen self-attention layers in the FPLM and PCA perform similar functions, which is not a coincidence. t is noteworthy that the degradation representation vectors, after patch processing, are still high-dimensional vectors with noisy multi-channels at each time step. In this situation, the FPLM with frozen self-attention layer parameters appears to decompose and project complex input patterns during gradient descent, suppressing noise and finding the intrinsic degradation behavior and its trajectory, thereby improving the signal-to-noise ratio of the input data pattern.

Additionally, we linearly increased the proportion of randomly initialized parameters from 10% to 100% and observed the experimental results. As shown in Figure 16-(a), as fewer parameters are frozen, i.e., less pre-trained knowledge is retained, the model's performance is severely compromised. This also indirectly proves the reason for selectively freezing the self-attention parameters from the pre-training stage while only fine-tuning the Normalization layers. This is because the pre-trained FPLM should be considered as a whole, and fine-tuning the self-attention layers would lead to a decline in the model's overall performance, as this operation disrupts the knowledge acquired during the pre-training stage.

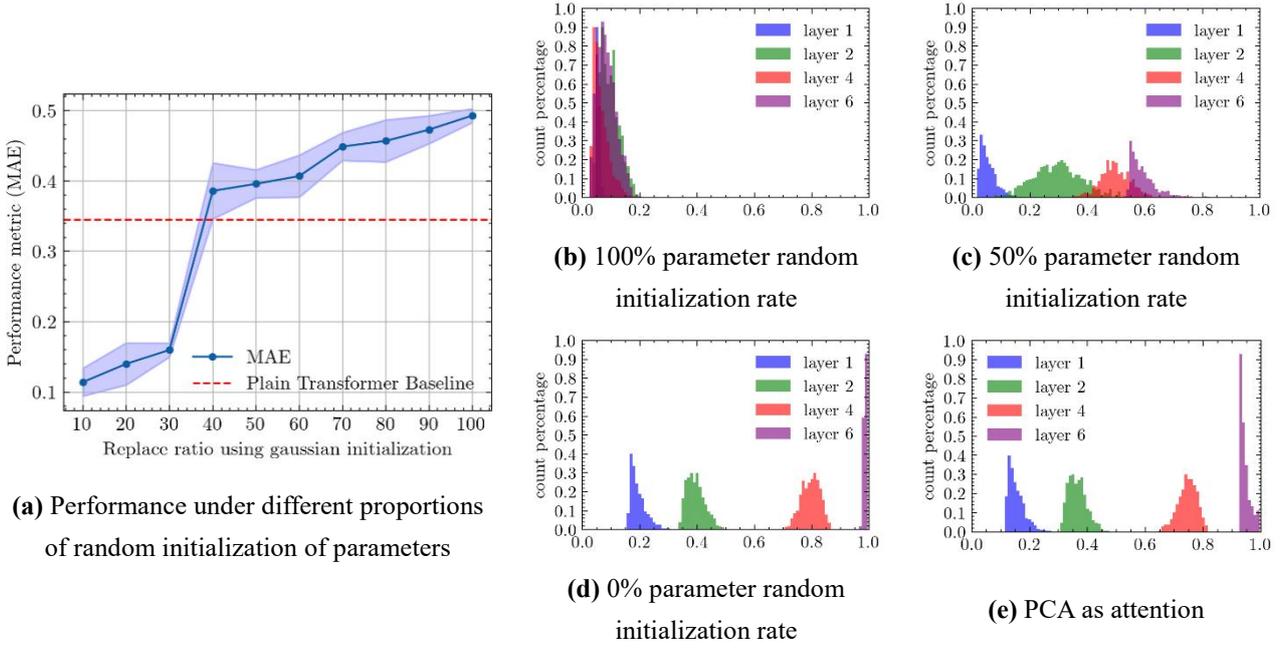

Figure 16 Statistical characteristics comparison of feature extraction behaviors between PCA and FPLM-HEL: **(a)** RUL prediction performance under different proportions of random initialization of parameters, evaluated by MAE metric; **(b)** Histogram of degradation feature distribution intervals when 0% of FPLM-HEL self-attention layer parameters are frozen, i.e., fully randomly initialized; **(c)** Histogram of degradation feature distribution intervals when 50% of FPLM-HEL self-attention layer parameters are frozen; **(d)** Histogram of degradation feature distribution intervals when 100% of FPLM-HEL self-attention layer parameters are frozen; **(e)** Histogram of degradation feature distribution intervals when PCA replaces self-attention layers

## 5. Conclusions

Accurate RUL prediction is pivotal for ensuring the reliability of bearings and minimizing unexpected equipment failures. Despite significant progress, existing data-driven deep learning methods face substantial challenges, including discrepancies in data distributions under real industrial conditions and limited generalization for long-term predictions. To tackle these issues, this work introduces LM4RUL, a robust framework that leverages pretrained large language models (LLMs) to enhance long-term RUL prediction. Key innovations include the Local Scale Perception Representation (LSPR), which effectively captures detailed degradation information by tokenizing vibration data, and the Hybrid Embedding Learning (HEL) component, which selectively fine-tunes model layers to retain critical predictive knowledge for modeling nonlinear degradation patterns. The two-stage fine-tuning strategy helps the framework adapt to data discrepancies and extend its applicability across varied conditions.

A common concern with industrial applications is the scalability of large models with limited data availability. Our research addresses this by demonstrating that LM4RUL not only outperforms smaller models in prediction accuracy but also maintains manageable inference costs, making it practical for real-world industrial applications. This advancement pushes the boundary of using large models effectively even in low-sample scenarios, emphasizing both their generalization and predictive capabilities.

There are, however, two areas for future enhancement. Firstly, the performance of our framework is inherently linked to both the quality of the pretrained LLMs and the effectiveness of the fine-tuning strategy. As the field of time series analysis continues to evolve, integrating advanced techniques like LoRA fine-tuning can potentially improve overall predictive accuracy. Secondly, although initial results suggest a promising advantage of large models for time series tasks, understanding the underlying mechanisms and enhancing model interpretability remain critical. Future work will focus on improving the interpretability of the LM4RUL framework, ensuring that its insights are transparent and actionable for industrial stakeholders.

# 6. Declaration of Competing Interest

The authors declare that they have no known competing financial interests or personal relationships that could have appeared to influence the work reported in this paper.

Oberlin, T., Meignen, S., & Perrier, V. (2014). The fourier-based synchrosqueezing transform. *2014 IEEE International Conference on Acoustics, Speech and Signal Processing (ICASSP)*, 315–319. https://doi.org/10.1109/ICASSP.2014.6853609

Ochella, S., Shafiee, M., & Dinmohammadi, F. (2022). Artificial intelligence in prognostics and health management of engineering systems. *Engineering Applications of Artificial Intelligence*, *108*, 104552. https://doi.org/10.1016/j.engappai.2021.104552

Qin, Y., Gan, F., Xia, B., Mi, D., & Zhang, L. (2024). Remaining useful life estimation of bearing via temporal convolutional networks enhanced by a gated convolutional unit. *Engineering Applications of Artificial Intelligence*, *133*, 108308. https://doi.org/10.1016/j.engappai.2024.108308

Shi, M., Ding, C., Shen, C., Huang, W., & Zhu, Z. (2024). Imbalanced class incremental learning system: A task incremental diagnosis method for imbalanced industrial streaming data. *Advanced Engineering Informatics*, *62*, 102832. https://doi.org/10.1016/j.aei.2024.102832

Su, X., Liu, H., Tao, L., Lu, C., & Suo, M. (2021). An end-to-end framework for remaining useful life prediction of rolling bearing based on feature pre-extraction mechanism and deep adaptive transformer model. *Computers & Industrial Engineering*, *161*, 107531. https://doi.org/10.1016/j.cie.2021.107531

Sun, Y., & Wang, Z. (2024). Remaining useful life prediction of rolling bearing via composite multiscale permutation entropy and Elman neural network. *Engineering Applications of Artificial Intelligence*, *135*, 108852. https://doi.org/10.1016/j.engappai.2024.108852

Wang, B., Lei, Y., Li, N., & Yan, T. (2019). Deep separable convolutional network for remaining useful life prediction of machinery. *Mechanical Systems and Signal Processing*, *134*, 106330. https://doi.org/10.1016/j.ymssp.2019.106330

Xiang, S., Qin, Y., Zhu, C., Wang, Y., & Chen, H. (2020). Long short-term memory neural network with weight amplification and its application into gear remaining useful life prediction. *Engineering Applications of*
41